\documentclass[aps,prd,twocolumn,superscriptaddress,showkeys,nofootinbib,amssymb]{revtex4-1}
\pdfoutput=1
\usepackage{graphicx}
\graphicspath{{figures/}}
\usepackage[utf8]{inputenc}

\begin{document}

\title{Probing Lepton Flavor Models at Future Neutrino Experiments}

\author{Mattias Blennow}
\email[]{emb@kth.se}
\affiliation{Department of Physics, School of Engineering Sciences, KTH Royal Institute of Technology, AlbaNova University Center, Roslagstullsbacken 21, SE--106~91 Stockholm, Sweden}
\affiliation{The Oskar Klein Centre for Cosmoparticle Physics, AlbaNova University Center, Roslagstullsbacken 21, SE--106 91 Stockholm, Sweden}
\affiliation{Departamento de F\'isica Te\'orica and Instituto de F\'{\i}sica Te\'orica, IFT-UAM/CSIC,
Universidad Aut\'onoma de Madrid, Cantoblanco, 28049, Madrid, Spain}\thanks{On leave of absence}

\author{Monojit Ghosh}
\email[]{manojit@kth.se}
\affiliation{Department of Physics, School of Engineering Sciences, KTH Royal Institute of Technology, AlbaNova University Center, Roslagstullsbacken 21, SE--106~91 Stockholm, Sweden}
\affiliation{The Oskar Klein Centre for Cosmoparticle Physics, AlbaNova University Center, Roslagstullsbacken 21, SE--106 91 Stockholm, Sweden}

\author{Tommy Ohlsson}
\email[]{tohlsson@kth.se}
\affiliation{Department of Physics, School of Engineering Sciences, KTH Royal Institute of Technology, AlbaNova University Center, Roslagstullsbacken 21, SE--106~91 Stockholm, Sweden}
\affiliation{The Oskar Klein Centre for Cosmoparticle Physics, AlbaNova University Center, Roslagstullsbacken 21, SE--106 91 Stockholm, Sweden}
\affiliation{University of Iceland, Science Institute, Dunhaga 3, IS--107 Reykjavik, Iceland}

\author{Arsenii Titov}
\email[]{arsenii.titov@pd.infn.it}
\affiliation{Dipartimento di Fisica e Astronomia ``G. Galilei'', Universit{\`a} degli Studi di Padova and INFN, Sezione di Padova, Via Francesco Marzolo 8, I--35131 Padova, Italy}


\begin{abstract}
Non-Abelian discrete symmetries provide an interesting opportunity to address the flavor puzzle in the lepton sector. However, the number of currently viable models based on such symmetries is rather large. High-precision measurements of the leptonic mixing parameters by future neutrino experiments, including ESSnuSB, T2HK, DUNE, and JUNO, will be crucial to test such models. We show that the complementarity among these experiments offers a powerful tool for narrowing down this broad class of lepton flavor models.
\end{abstract}

\maketitle

\section{Introduction}
\label{sec:intro}
%
In the Standard Model (SM) of particles and their interactions, 
quarks and leptons come in three generations or families.
This number as well as the values of 
fermion masses remain one of the main puzzles in particle physics.
We also do not know why the mixing patterns 
in the quark and lepton sectors are so different, 
and whether their structures point to any organizing principle or not. 
All these questions form the so-called flavor problem.

Lepton mixing featuring two large and one small mixing angles 
may originate from non-Abelian discrete flavor symmetries. 
Such a possibility has been widely explored over the past decades 
(see Refs.~\cite{Altarelli:2010gt,Ishimori:2010au,King:2013eh,Petcov:2017ggy,Feruglio:2019ktm} for reviews). 
Models based on such symmetries often make
predictions for the Dirac CP-violating phase $\delta_\mathrm{CP}$, 
which starts to be experimentally constrained~\cite{Abe:2019vii,Acero:2019ksn}.
An attractive feature of the discrete symmetry approach to lepton flavor 
is that its predictions can be tested at current and future neutrino 
experiments (see, e.g., Refs.~\cite{Antusch:2007rk,Ballett:2013wya,Petcov:2014laa,Girardi:2014faa,Ballett:2014dua,Girardi:2015vha,Agarwalla:2017wct,Petcov:2018snn,Blennow:2020snb}).

Quark mixing being small does not seem to favor 
non-Abelian discrete symmetries, and usually, 
the quark and lepton sectors are treated separately. 
Still, rather numerous attempts have been made to describe both sectors 
with discrete flavor symmetries as, e.g., 
$A_4$ ~\cite{Babu:2002dz,He:2006dk,Ma:2006sk,King:2006np} 
or $T'$~\cite{Feruglio:2007uu,Chen:2007afa,Frampton:2007et}, 
the latter being the double covering of the former.
Finding a unified solution 
to the flavor problem in both sectors is a formidable task 
and we will not try to address it in the present article. 
In what follows, we concentrate 
on the lepton sector alone.

In Ref.~\cite{Blennow:2020snb}, we considered a class of models 
based on the $A_4$, $S_4$, and $A_5$ finite groups~\cite{Feruglio:2012cw,Li:2015jxa,DiIura:2015kfa,Ballett:2015wia,Girardi:2015rwa}. 
These groups are minimal admitting a three-dimensional irreducible representation 
to which three lepton generations are assigned.
We confronted  the predictions of 18 models with current global 
neutrino oscillation data~\cite{Esteban:2018azc,NuFITv41} 
(see Refs.~\cite{Capozzi:2018ubv,deSalas:2020pgw} for alternative global analyses)
and found that ten models survive at $3\sigma$. 
We further explored the potential of the proposed ESSnuSB 
long-baseline (LBL) neutrino oscillation experiment~\cite{Baussan:2013zcy,Wildner:2015yaa} 
to discriminate among and exclude (under certain assumptions) these models. 
In this work, we address these ten lepton flavor models with other future LBL 
experiments, T2HK~\cite{Abe:2015zbg} and DUNE~\cite{Acciarri:2015uup,Abi:2020evt}, 
as well as with the medium-baseline reactor neutrino experiment JUNO~\cite{An:2015jdp,Djurcic:2015vqa}. 
We show that the complementarity among these experiments 
provides an effective way to constrain this class of models. 

The article is organized as follows. 
In Sec.~\ref{sec:models}, we introduce a set of lepton flavor models 
based on non-Abelian discrete symmetries 
and review their compatibility with the global neutrino oscillation data. 
Then, in Sec.~\ref{sec:experiments}, we describe the relevant experimental setups 
and provide the simulation details along with the statistical method used. 
Next, in Sec.~\ref{sec:results}, we present the results of our statistical analysis. 
Finally, in Sec.~\ref{sec:conclusions}, we summarize and draw our conclusions.

\section{Lepton Flavor Models}
\label{sec:models}
%
A non-Abelian discrete flavor symmetry $G_f$ can be consistently combined 
with a generalized CP symmetry~\cite{Feruglio:2012cw,Holthausen:2012dk}. 
Breaking the full symmetry group to a residual symmetry
$G_e = Z_k$, $k>2$ or $Z_m \times Z_n$, $m,n \geq 2$  
in the charged lepton sector and a remnant symmetry 
$G_\nu = Z_2 \times \text{CP}$ in the neutrino sector 
leads to a leptonic mixing matrix $U_\mathrm{PMNS}$ 
that depends on a single free angle. 
All leptonic mixing parameters are functions of this angle 
and therefore highly correlated.
In the cases of $G_f = S_4$ and $A_5$, the corresponding mixing patterns 
were derived in Refs.~\cite{Feruglio:2012cw} and \cite{Li:2015jxa,DiIura:2015kfa,Ballett:2015wia}, respectively. 
All of them lead to sharp predictions for the leptonic mixing parameters 
$\theta_{12}$, $\theta_{13}$, $\theta_{23}$, and $\delta_\mathrm{CP}$.
If we relax the assumption of CP symmetry and break $G_f$ to either
$G_e = Z_2$ and $G_\nu = Z_k$, $k>2$ or $Z_m \times Z_n$, $m,n \geq 2$,
or 
$G_e = Z_k$, $k>2$ or $Z_m \times Z_n$, $m,n \geq 2$ and $G_\nu = Z_2$, 
$U_\mathrm{PMNS}$ depends on two free parameters. 
Such possibilities were investigated in Ref.~\cite{Girardi:2015rwa} 
for $G_f = A_4$, $S_4$, and $A_5$. 

In Ref.~\cite{Blennow:2020snb}, we demonstrated that
out of the eleven (seven) one-(two-)parameter models, 
five (five) are compatible with the present global data at $3\sigma$. 
We summarize them along with their predictions in Table~\ref{tab:models}.
Among the five two-parameter models, 
one leads to a sharp prediction for $\theta_{23}$ 
and a correlation between $\theta_{12}$ and $\delta_\mathrm{CP}$, 
whereas four give distinct predictions for $\theta_{12}$ 
and yield correlations between $\theta_{23}$ and $\delta_\mathrm{CP}$. 
The above-mentioned correlations involving $\delta_\mathrm{CP}$ 
are denoted by $f_k(\theta_{ij})$ in the table. 
The $\chi^2$ function, the minimum of which we quote in the last column, 
is defined according to Eqs.~(3.1) and (3.2) of Ref.~\cite{Blennow:2020snb}. 
\begin{table}[t]
\renewcommand*{\arraystretch}{1.2}
\centering
\begin{tabular}{|c|c|c|ccc|c|}
\hline
Model & Case~[Ref.] & Group & $\sin^2 \theta_{12}$ & $\sin^2 \theta_{23}$ & $\delta_{\rm CP}$ & $\chi^2_\mathrm{min}$ \\
\hline
\hline
1.1 & VII-b~\cite{Li:2015jxa} & $A_5 \rtimes \text{CP}$ & 0.331 & 0.523 & $180^\circ$ & 5.37 \\
1.2 & III~\cite{Li:2015jxa} & $A_5 \rtimes \text{CP}$ & 0.283 & 0.593 & $180^\circ$ & 5.97 \\
1.3 & IV~\cite{Feruglio:2012cw} & $S_4 \rtimes \text{CP}$ & 0.318 & $1/2$ & $\pm90^\circ$ & 7.28 \\
1.4 & II~\cite{Feruglio:2012cw} & $S_4 \rtimes \text{CP}$ & 0.341 & 0.606 & $180^\circ$ & 8.91 \\
1.5 & IV~\cite{Li:2015jxa} & $A_5 \rtimes \text{CP}$ & 0.283 & $1/2$ & $\pm90^\circ$ & 11.3 \\
\hline
\hline
2.1 & A1~\cite{Girardi:2015rwa} & $A_5$ & | & 0.554 & $f_{1}(\theta_{12})$ & 0.151  \\
2.2 & B2~\cite{Girardi:2015rwa} & $S_4$ & 0.318 & | & $f_{2}(\theta_{23})$ & 0.386   \\
2.3 & B2~\cite{Girardi:2015rwa} & $A_5$ & 0.330 & | & $f_{3}(\theta_{23})$ & 2.49  \\
2.4 & B1~\cite{Girardi:2015rwa} & $A_5$ & 0.283 & | & $f_{4}(\theta_{23})$ & 4.40  \\
2.5 & B1~\cite{Girardi:2015rwa} & $A_4/S_4/A_5$ & 0.341 & | & $f_{5}(\theta_{23})$ &  5.67  \\
\hline
\end{tabular}
\caption{Predictions of the one- and two-parameter models compatible with the global data at $3\sigma$. A rational number means an exact prediction for the corresponding mixing parameter, whereas a decimal value implies that the mixing parameter lies in a very narrow interval around this value.}
\label{tab:models}
\end{table}
%

Finally, let us note that the predictions for the leptonic mixing parameters
are subject to renormalization group (RG) corrections. 
Such corrections depend on a model describing physics 
between a high-energy scale at which the predictions are derived 
and a low-energy scale at which the neutrino oscillation parameters 
are measured. In the SM and its minimal supersymmetric extension (MSSM), 
augmented with either the Weinberg dimension-5 operator 
or heavy electroweak singlet neutrinos to generate the small neutrino masses, 
the RG evolution of the leptonic mixing parameters is known~\cite{Antusch:2003kp,Antusch:2005gp} (for a review, see Ref.~\cite{Ohlsson:2013xva}). 
In the context of lepton flavor models with discrete symmetries, 
RG corrections have been studied in Refs.~\cite{Ballett:2014dua,Zhang:2016djh,Zhang:2016png,Gehrlein:2016fms}. 
It has been found that in the SM the RG effects are negligible, 
whereas in the MSSM they can be sizable if
the smallest neutrino mass is larger than about $0.01$~eV 
and $\tan\beta \gtrsim 30$~\cite{Gehrlein:2016fms}. 
In the latter case, the predictions realized at a high-energy scale 
are generally washed out at low energies. 
Nevertheless, if the predictions are shifted rather than washed out 
upon RG evolution, 
the methodology that we present in the next section can still be applied.

\section{Future Neutrino Experiments and Statistical Method}
\label{sec:experiments}
%
We simulate all experimental setups under consideration 
using the GLoBES software~\cite{Huber:2004ka,Huber:2007ji}. 
For ESSnuSB, we use the same configuration as in Refs.~\cite{Ghosh:2019sfi,Blennow:2019bvl,Ghosh:2019zvl,Blennow:2020snb}. 
We consider a 1~Mt water-Cherenkov detector located 540~km 
from the neutrino source capable of delivering 
$2.7 \times 10^{23}$ protons on target (POT) per year for 10~years 
with a beam power of 5~MW. 
We also consider a functionally identical 0.1~kt near detector 
500~m from the source. 
The systematic uncertainties between the near and far detectors 
are correlated and given in Table~10 of Ref.~\cite{Blennow:2020snb}. 
For T2HK, we use the configuration given in Ref.~\cite{Abe:2016ero}. 
We consider two water-Cherenkov detectors of 187~kt each 
located 295~km from the source having a beam power of 1.3~MW 
with a total exposure of $27 \times 10^{21}$ POT, 
corresponding to 10~years of running.
For DUNE, we use the official GLoBES files 
of the DUNE technical design report~\cite{Abi:2020evt}. 
A 40~kt liquid argon time-projection chamber detector 
is placed 1300~km from the source having a power of 1.2~MW 
delivering $1.1 \times 10^{21}$ POT per year with a running time of 7~years.
For JUNO, we consider the configuration used in Refs.~\cite{Forero:2017vrg,Huber:2019frh} and based on details given in Ref.~\cite{An:2015jdp}. 
We consider a 20~kt liquid-scintillator detector 
with an energy resolution of $3~\%/\sqrt{E}$ 
located 53 km from the nuclear reactor source 
having a total thermal power of 36~${\rm GW_{th}}$. 
We also consider a near detector with resolution $1.7~\%/\sqrt{E}$ 
located 30~m from a reactor core 
having a thermal power of 4.6~${\rm GW_{th}}$. 
For T2HK, DUNE, and JUNO, the systematic errors are adopted 
from Refs.~\cite{Abe:2016ero}, \cite{Abi:2020evt}, and \cite{An:2015jdp}, respectively.
For all LBL experiments, we assume an equal amount of POT 
in the neutrino and antineutrino modes. 
For JUNO, we consider a total running time of 6 years. 

The statistical treatment of our results is based on the GLoBES $\chi^2$ function 
for the simulated experiments, which is essentially given by 
the summation of the individual Poisson log-likelihoods
\begin{equation}
\chi_{G}^2(\theta) = \sum_i \left[\bar D_i(\theta) - D_i + D_i \ln\left(\frac{D_i}{\bar D_i(\theta)}\right)\right] \,,
\end{equation}
%
where $D_i$ is the number of observed events in bin $i$ 
and $\bar D_i(\theta)$ the theoretically expected number of events 
for some given parameter set $\theta$. 
We let $D_i$ be given by the Asimov data~\cite{Cowan:2010js} 
predicted by a set of true leptonic mixing parameters 
depending on the assumptions on the parameter values in the true model. 
To the GLoBES $\chi^2$ function, we add a Gaussian prior 
$\chi^2_{\rm pr}(\theta)$ on $\theta_{12}$, $\theta_{23}$, and $\theta_{13}$ 
based on the global data as described in Eq.~(3.1) of Ref.~\cite{Blennow:2020snb}. We then compute the minimum total $\chi^2$ function for a model as
\begin{equation}
\chi^2_{\rm min} = \min_\theta [\chi^2_G(\theta) + \chi^2_{\rm pr}(\theta)]
\end{equation}
%
and define $\Delta\chi^2 = \chi^2_{\rm min} - \chi^2_0$, 
where $\chi^2_0$ is the minimum $\chi^2$ in a model 
where all leptonic mixing parameters are allowed to vary freely 
rather than being constrained by a flavor model. 
For the cases where Wilk's theorem~\cite{Wilks:1938dza} 
can be assumed to hold, $\Delta \chi^2$ is expected to be 
$\chi^2$-distributed with $N-n$ degrees of freedom (d.o.f.), 
where $n$ is the number of parameters of the model under scrutiny 
and $N$ the number of parameters relevant to the setup 
in the case where all leptonic mixing parameters vary freely.

\section{Results of Statistical Analysis}
\label{sec:results}
\begin{figure*}[t]
\centering
\includegraphics[width=0.31\textwidth]{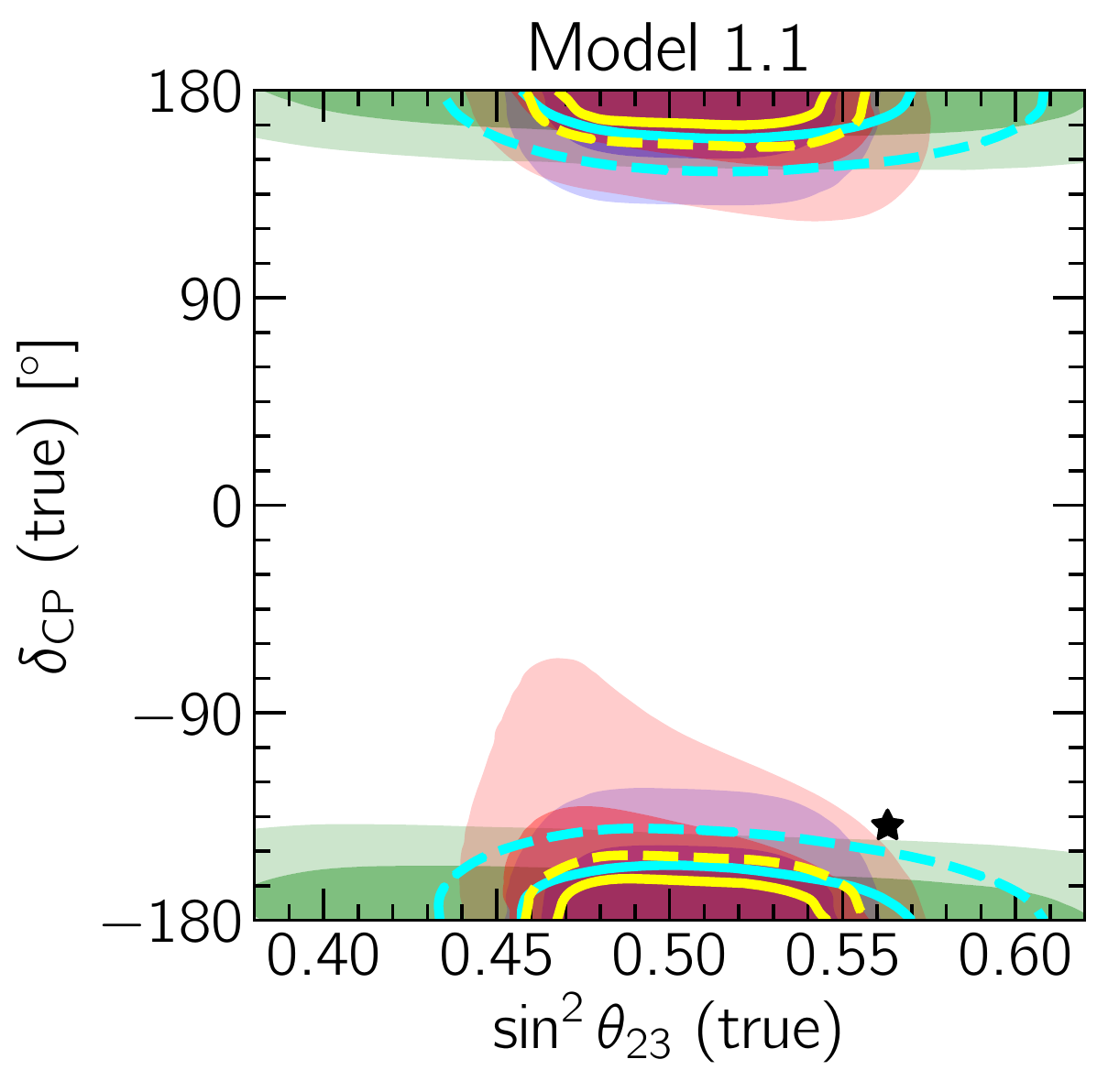}
\hspace{0.02\textwidth}
\includegraphics[width=0.31\textwidth]{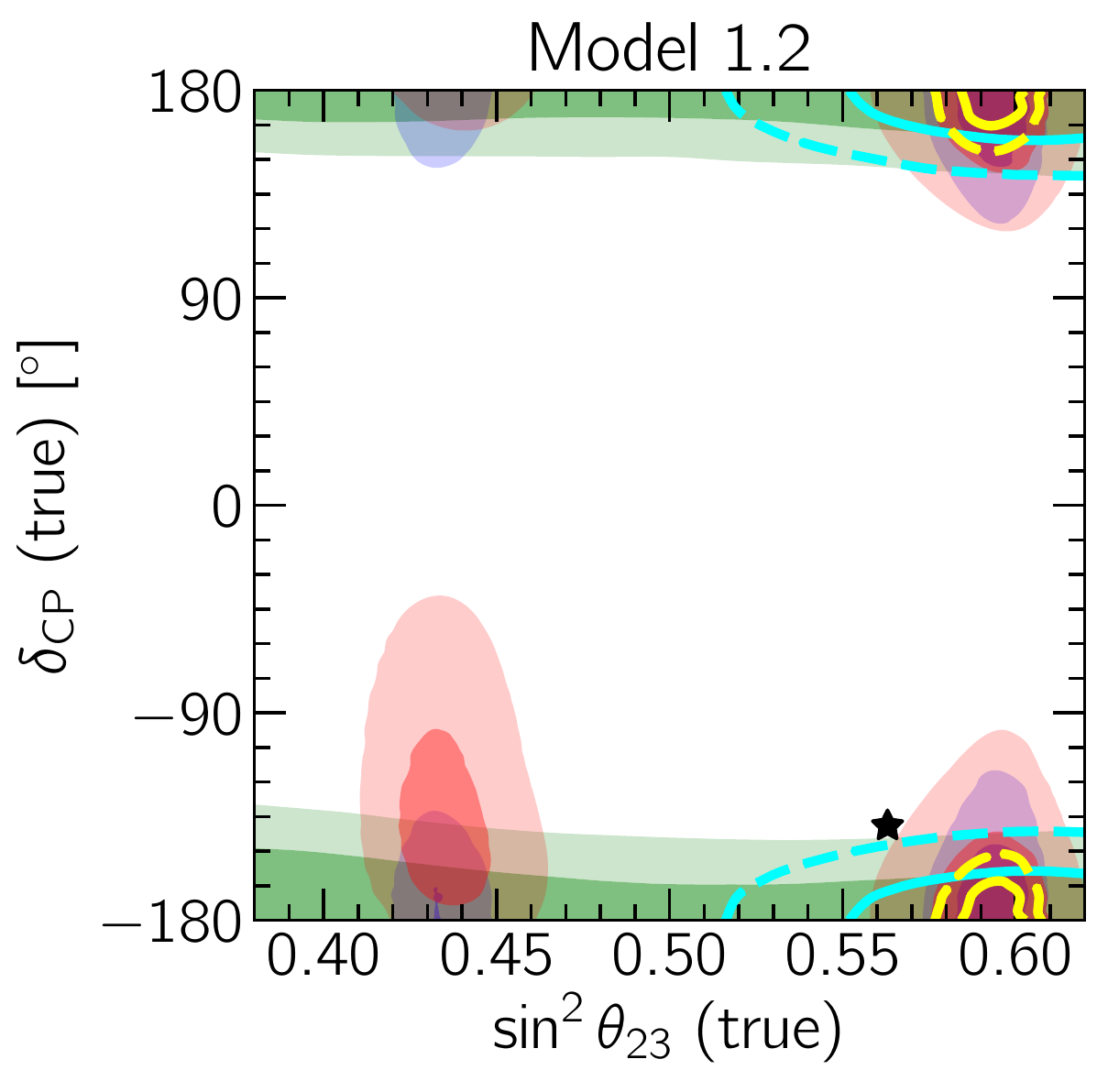}
\hspace{0.02\textwidth}
\includegraphics[width=0.31\textwidth]{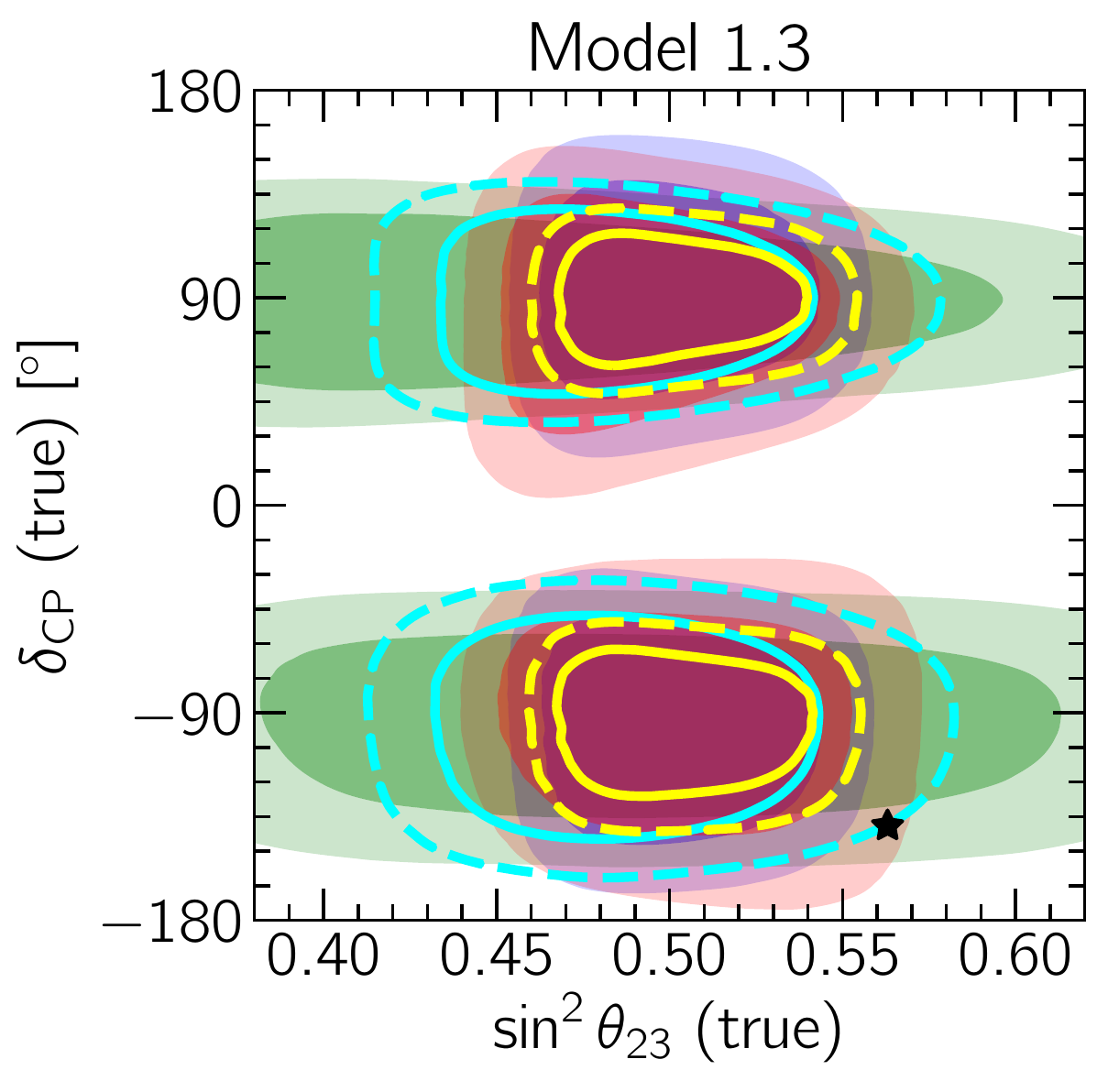}
\\[0.02\textwidth]
\includegraphics[width=0.31\textwidth]{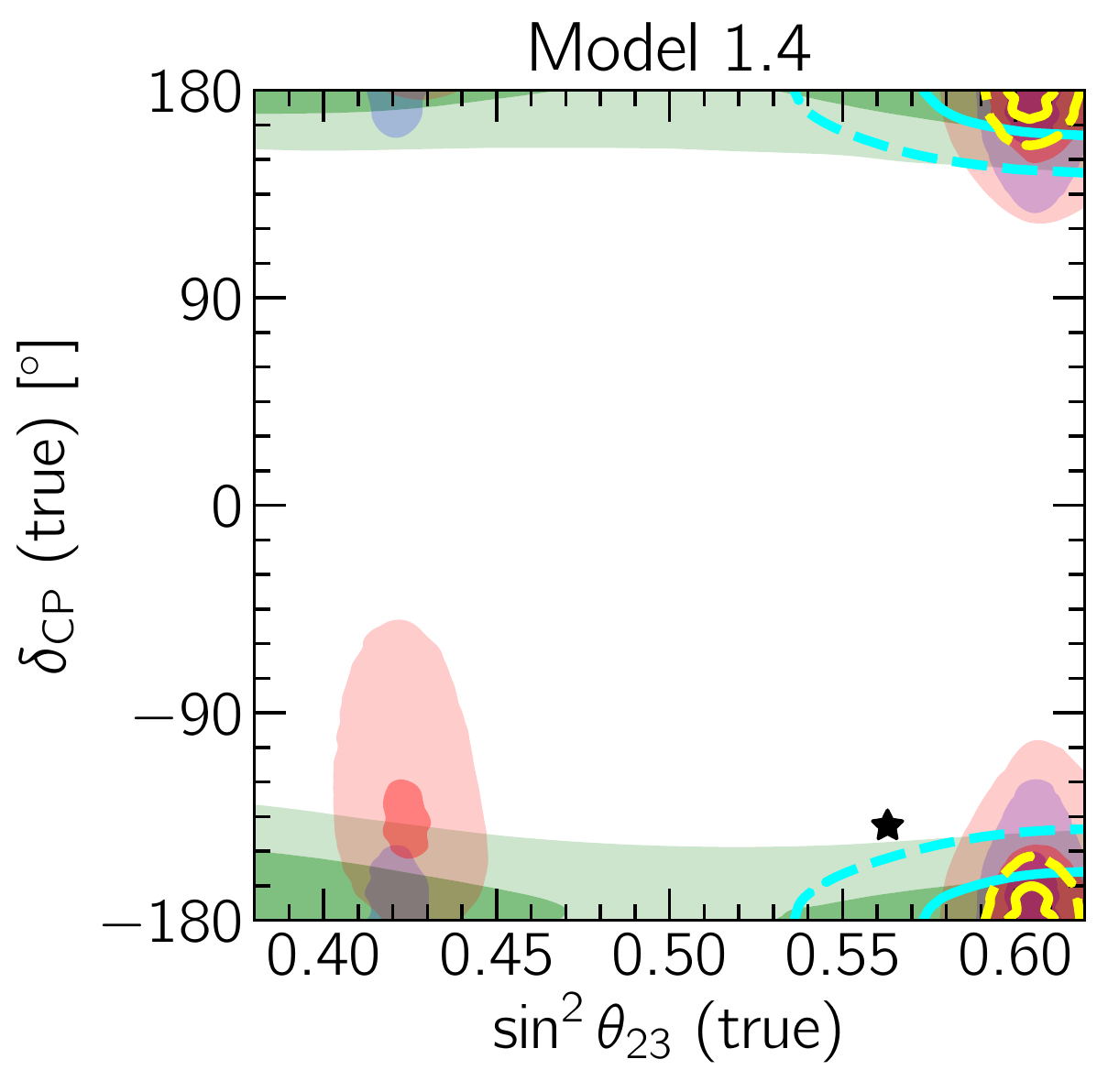}
\hspace{0.02\textwidth}
\includegraphics[width=0.31\textwidth]{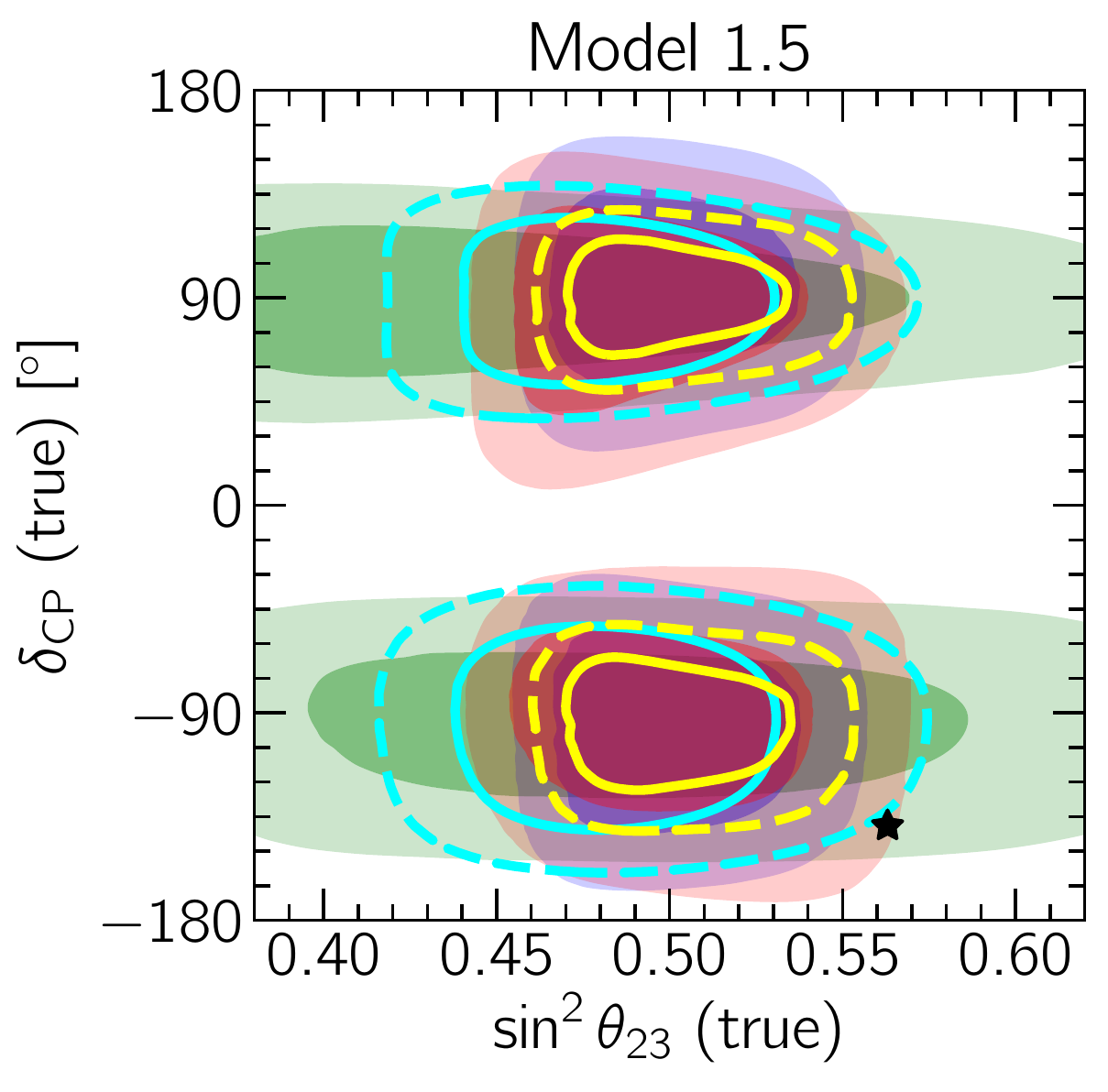}
\hspace{0.02\textwidth}
\includegraphics[width=0.31\textwidth]{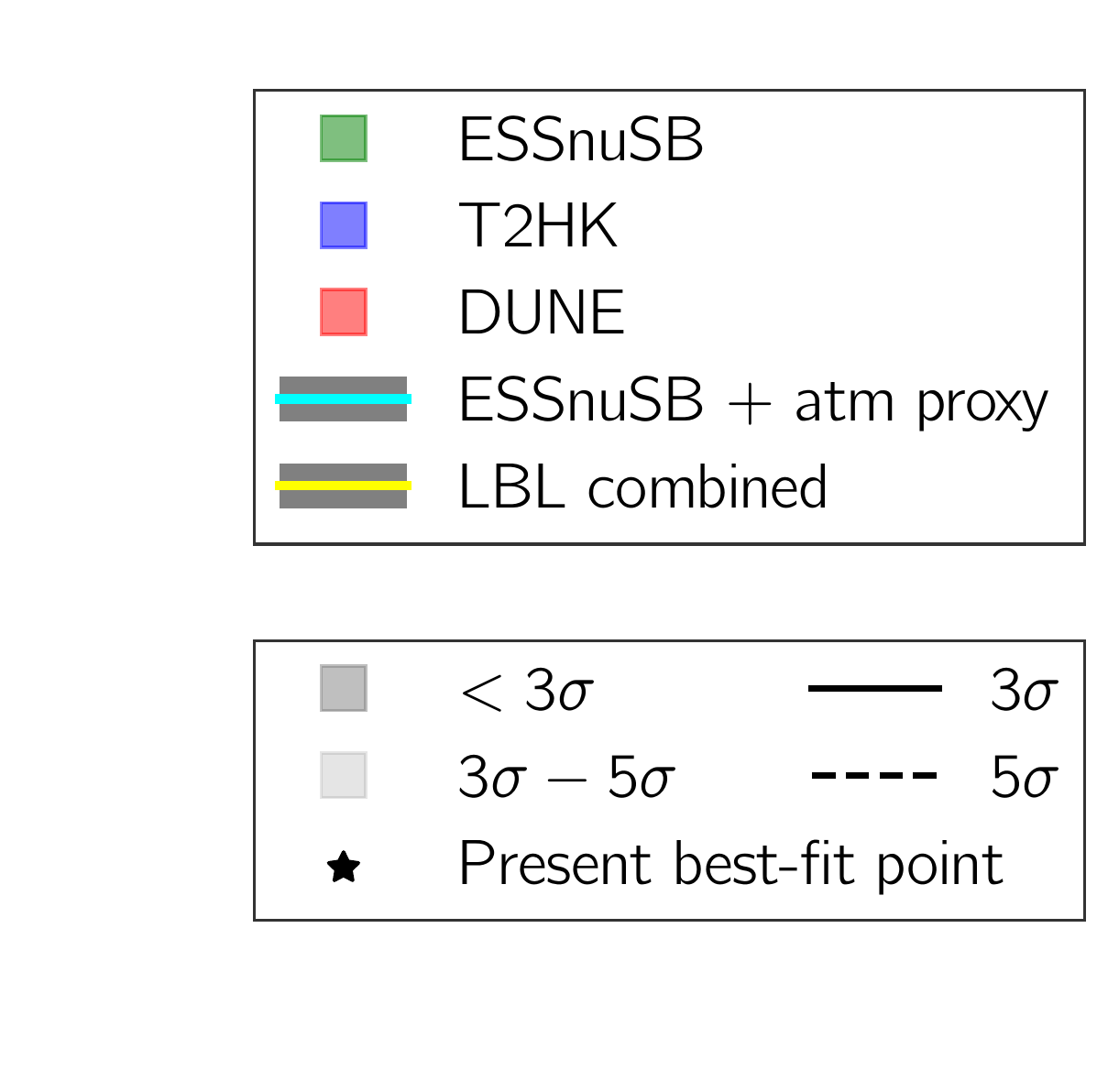}
\caption{Compatibility of one-parameter models with any potentially true values of $\sin^2\theta_{23}$ and $\delta_\mathrm{CP}$ in the context of ESSnuSB, T2HK, DUNE, and their combination denoted as ``LBL combined''. The ``ESSnuSB + atm proxy'' mimics the addition of the atmospheric neutrino data sample that the far detector of ESSnuSB would collect. The filled regions and contours correspond to the indicated number of $\sigma$'s for 3 d.o.f.}
\label{fig:one-param-t23delta}
\end{figure*}

\begin{figure*}[t]
\centering
\includegraphics[width=0.31\textwidth]{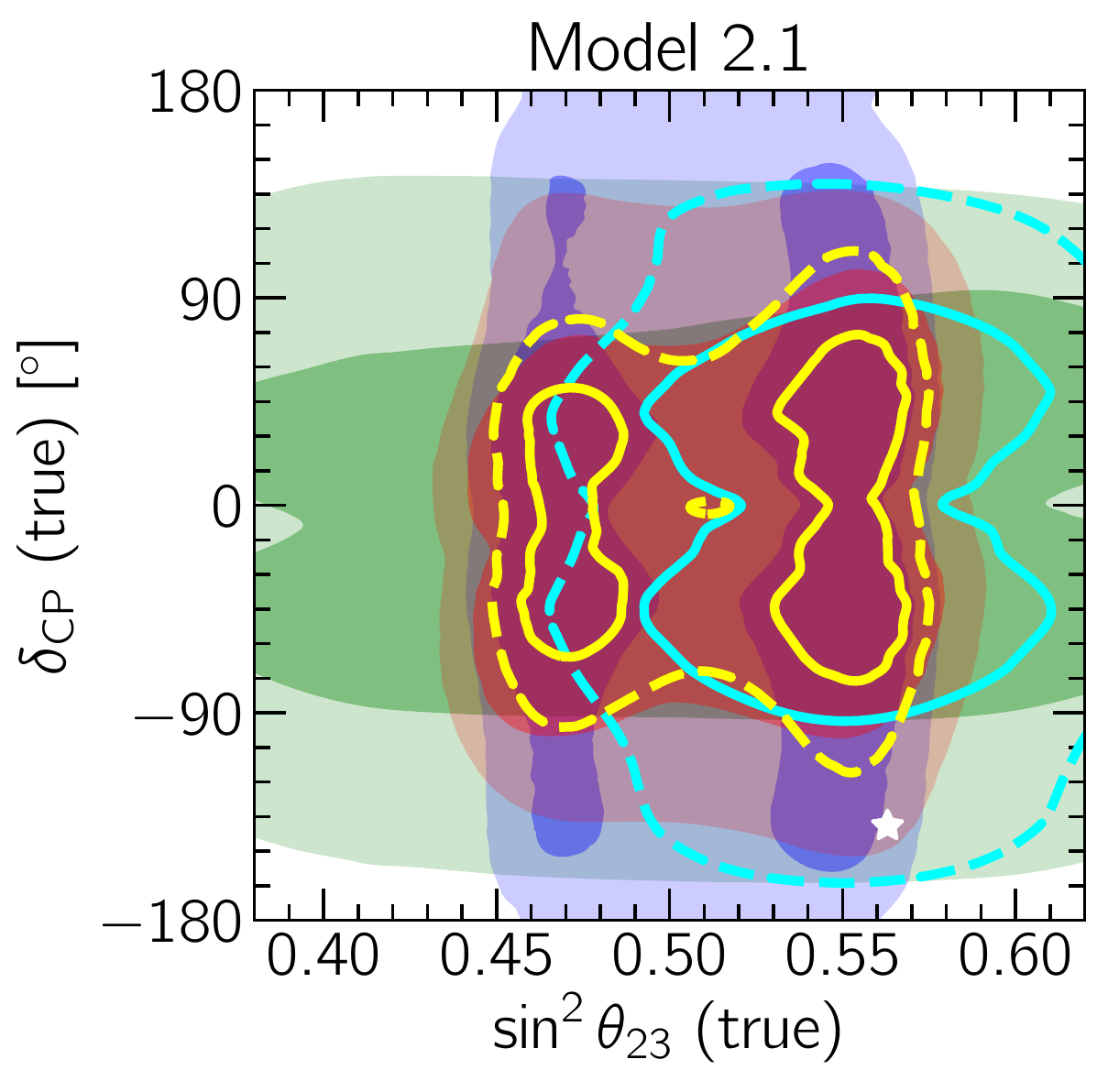}
\hspace{0.02\textwidth}
\includegraphics[width=0.31\textwidth]{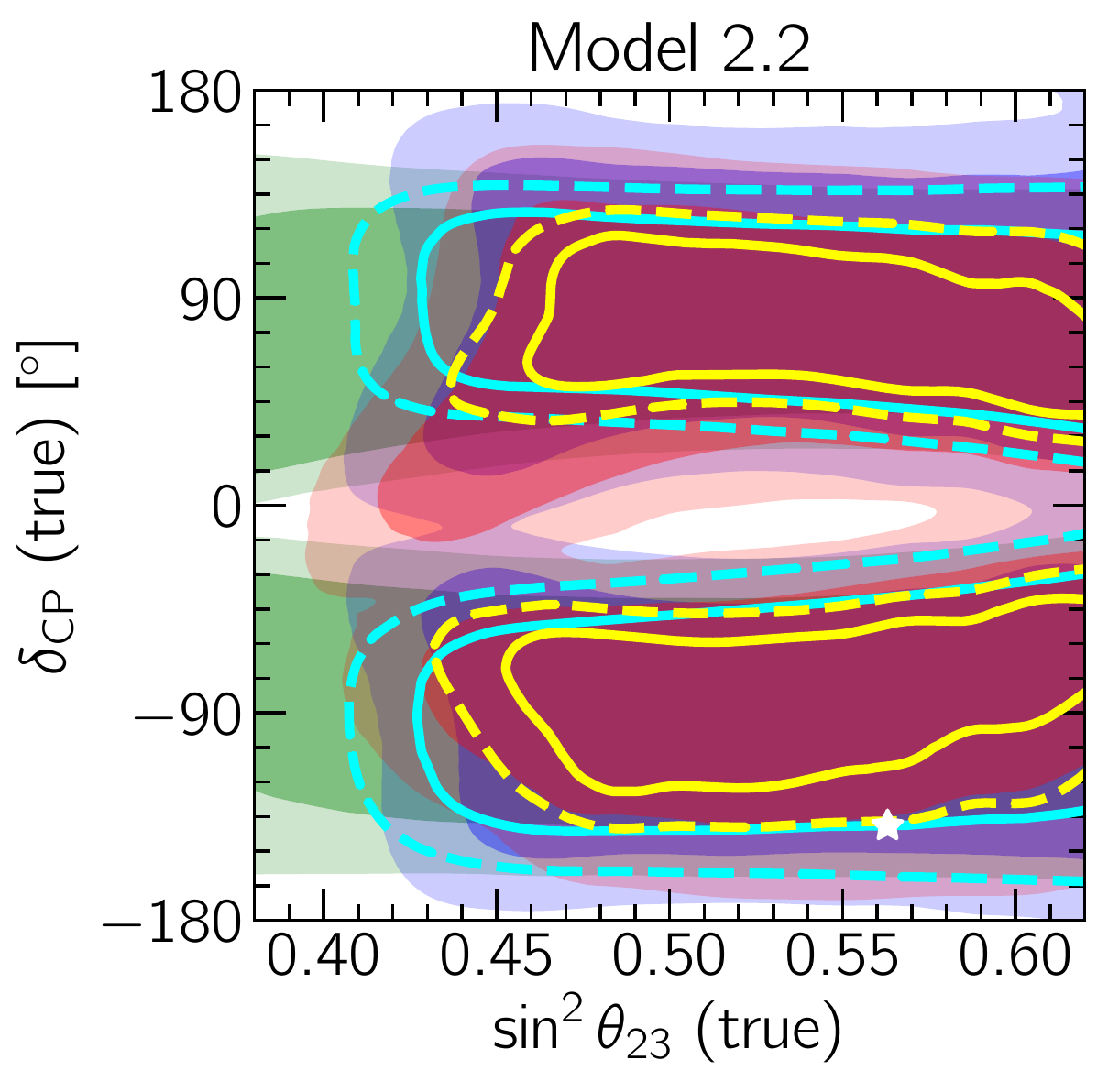}
\hspace{0.02\textwidth}
\includegraphics[width=0.31\textwidth]{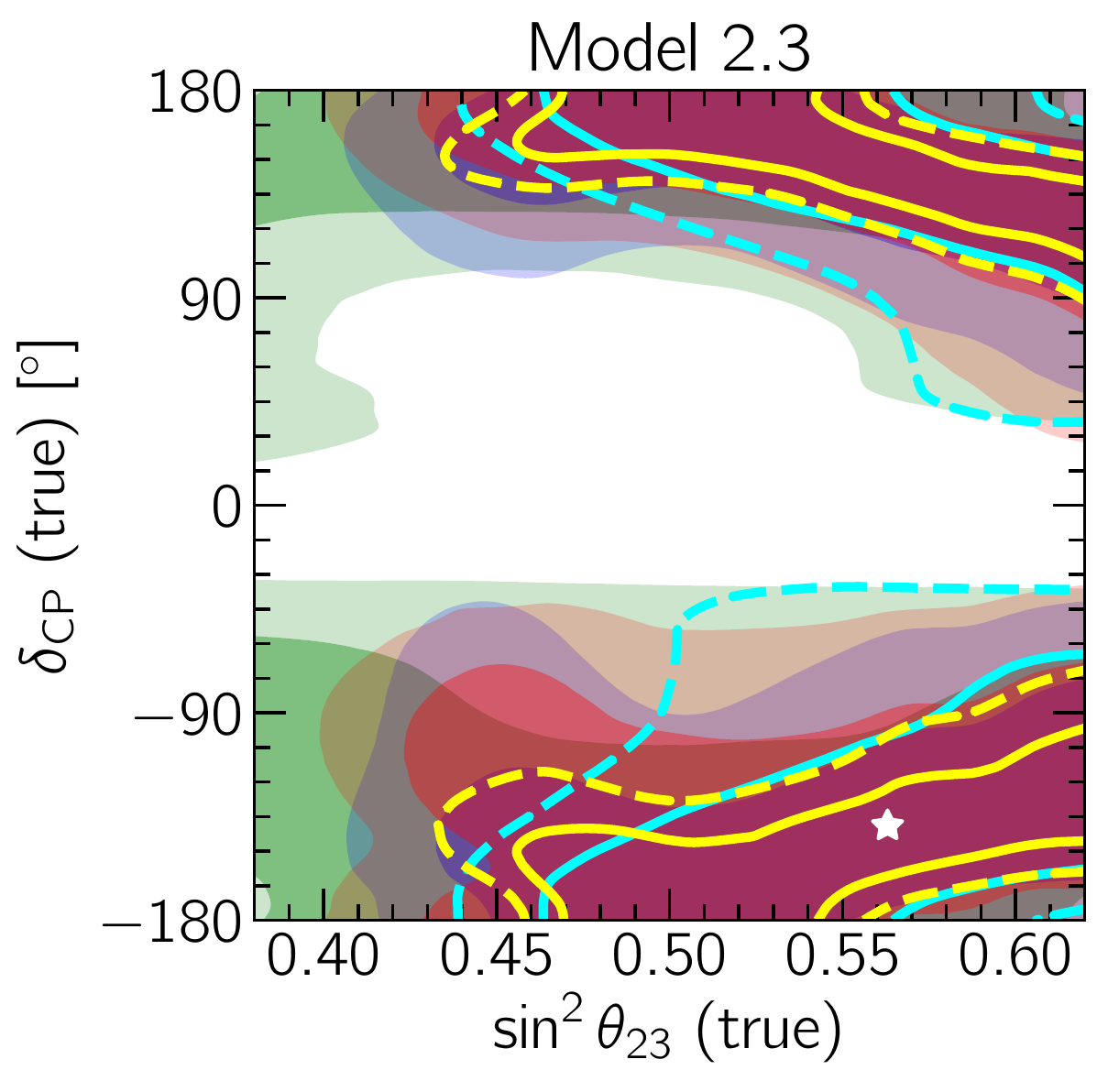}
\\[0.02\textwidth]
\includegraphics[width=0.31\textwidth]{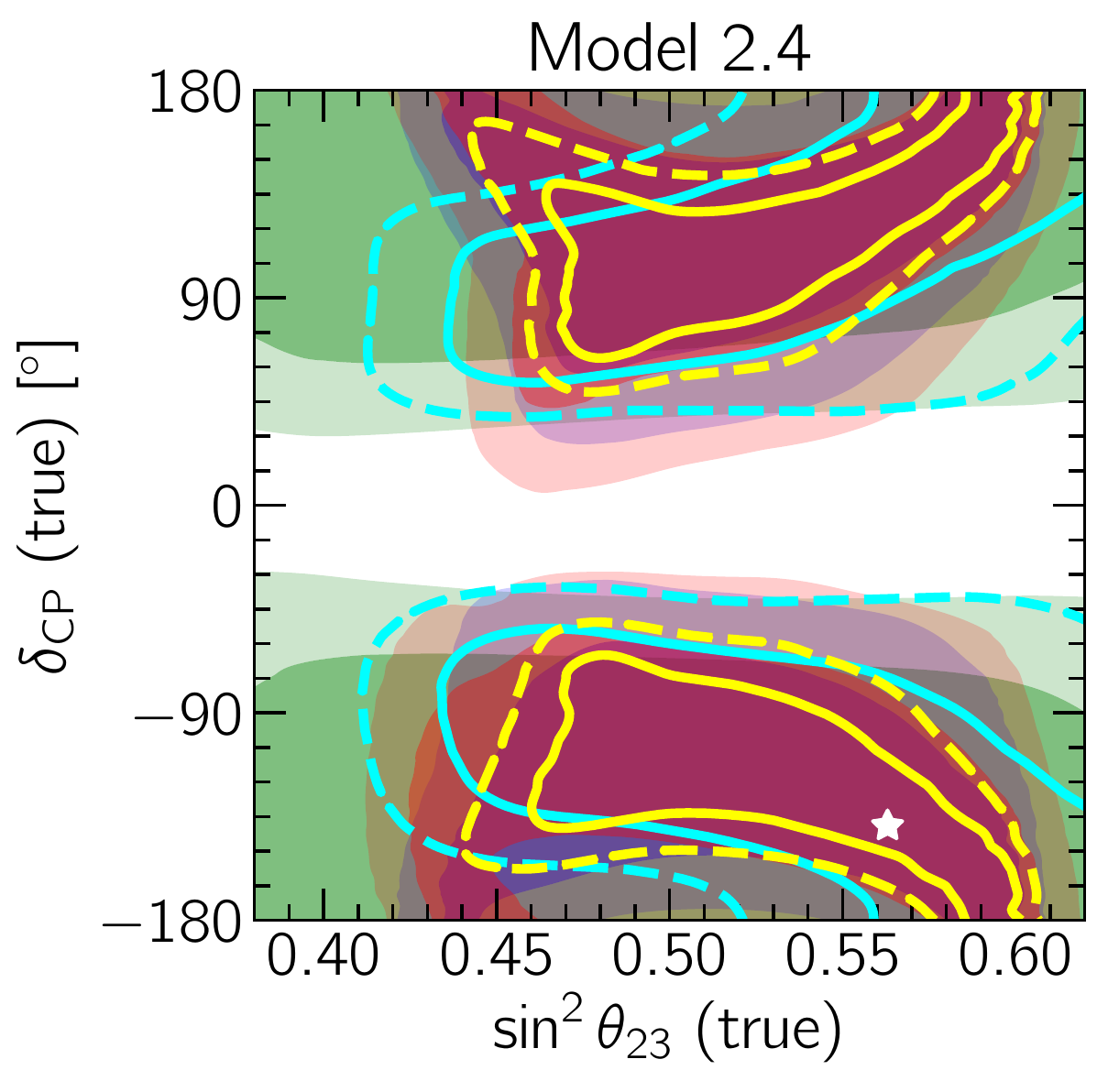}
\hspace{0.02\textwidth}
\includegraphics[width=0.31\textwidth]{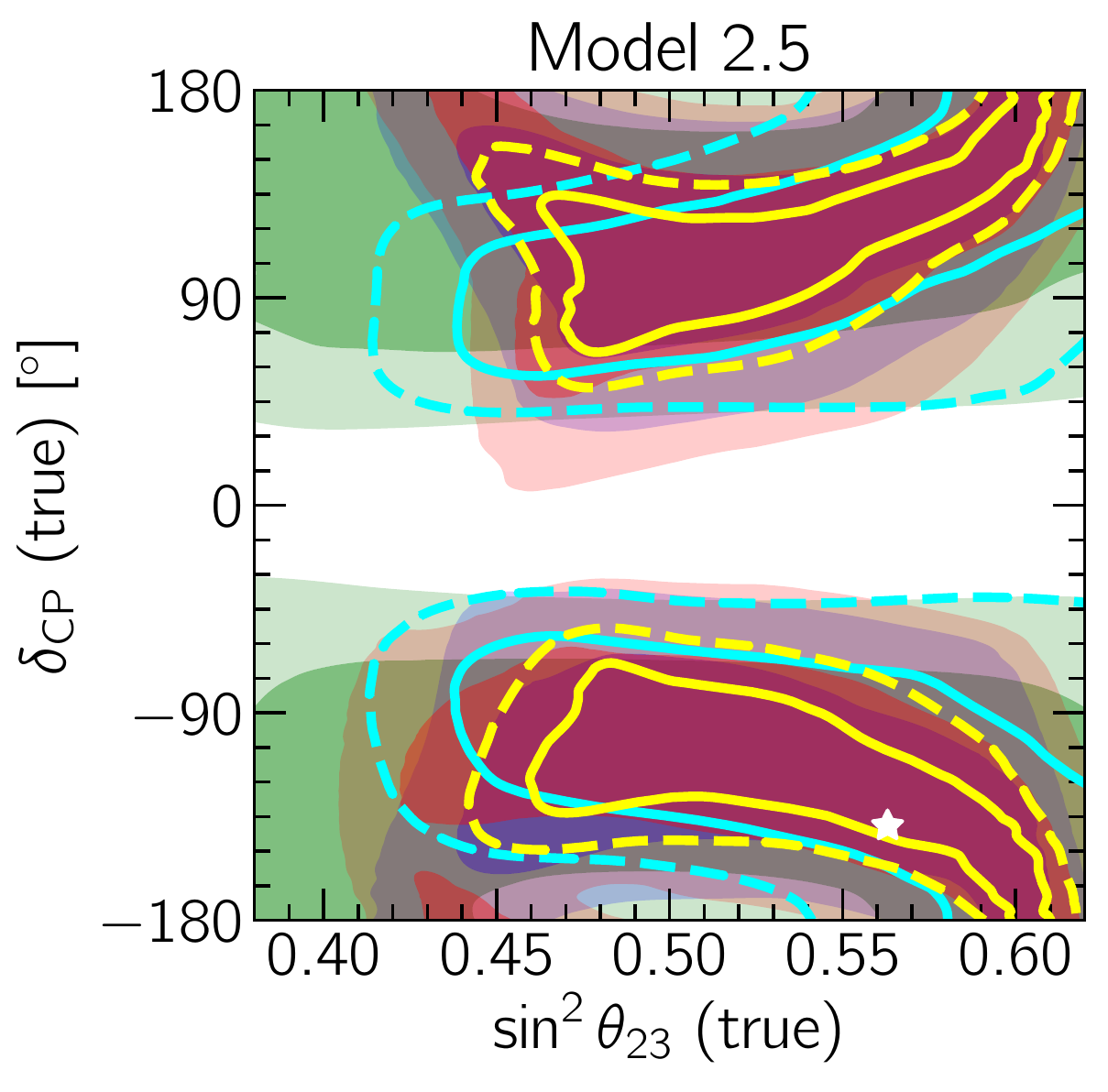}
\hspace{0.02\textwidth}
\includegraphics[width=0.31\textwidth]{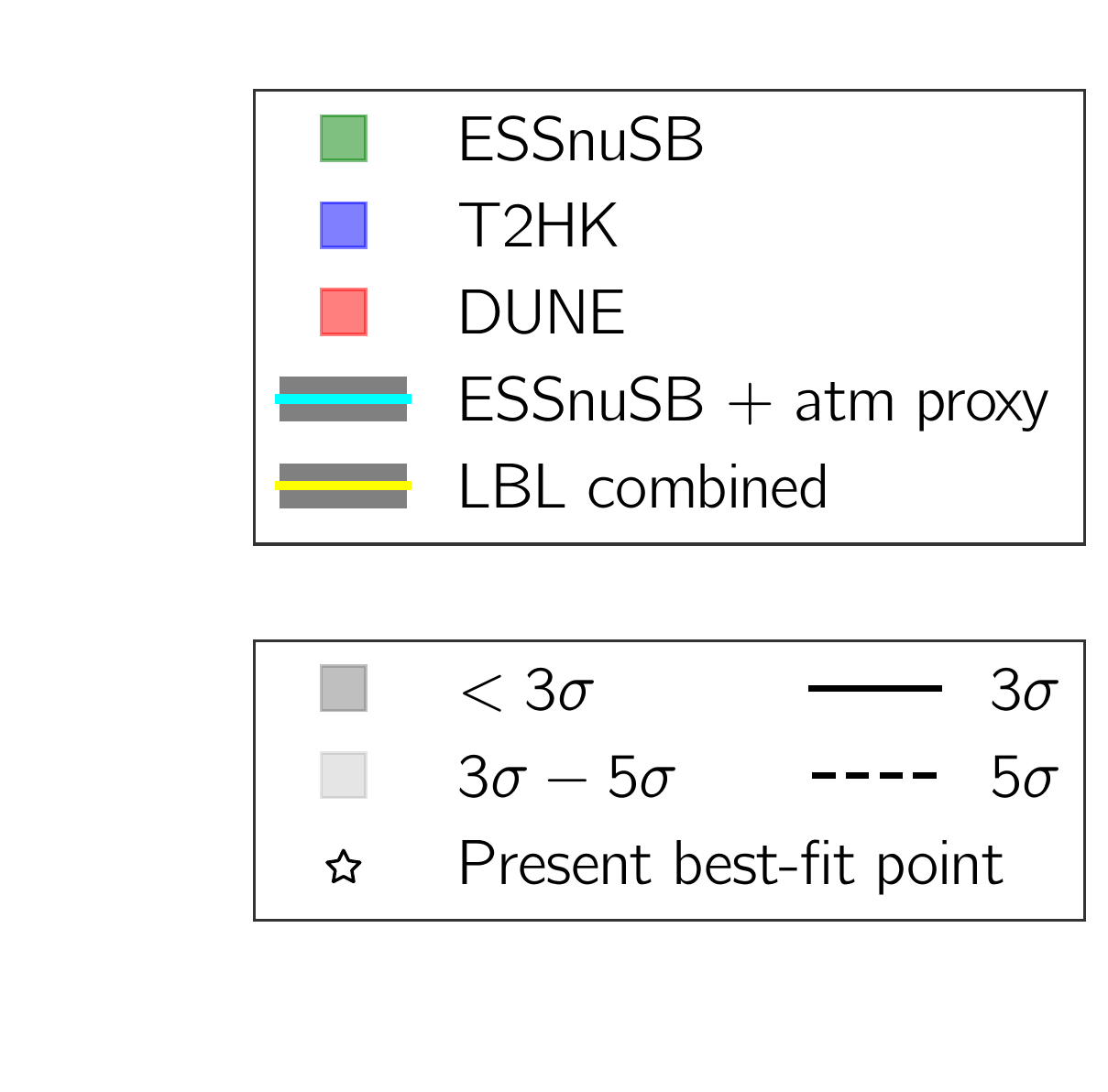}
\caption{Compatibility of two-parameter models with any potentially true values of $\sin^2\theta_{23}$ and $\delta_\mathrm{CP}$ in the context of ESSnuSB, T2HK, DUNE, and their combination denoted as ``LBL combined''. The ``ESSnuSB + atm proxy'' mimics the addition of the atmospheric neutrino data sample that the far detector of ESSnuSB would collect. The filled regions and contours correspond to the indicated number of $\sigma$'s for 2 d.o.f.}
\label{fig:two-param-t23delta}
\end{figure*}

%
In Figs.~\ref{fig:one-param-t23delta} and \ref{fig:two-param-t23delta}, 
we present the capability of ESSnuSB, T2HK, and DUNE 
as well as the combination of the three setups (LBL combined) 
to exclude the one- and two-parameter models 
in the $\sin^2 \theta_{23}$(true)--$\delta_{\rm CP}$(true) plane.
For the other neutrino oscillation parameters, we assume the true values: 
$\sin^2 \theta_{12} = 0.310$, $\sin^2 \theta_{13} = 0.02237$, 
$\Delta m_{21}^2 = 7.39 \times 10^{-5} \, {\rm eV}^2$, 
and $\Delta m_{31}^2 = 2.528 \times 10^{-3} \, {\rm eV}^2$, 
i.e., normal ordering (NO) of neutrino masses.
The regions for ESSnuSB were derived in Ref.~\cite{Blennow:2020snb}.
If the true values of $\sin^2\theta_{23}$ and $\delta_\mathrm{CP}$
fall inside the regions, the test model is compatible with the Asimov data 
at the shown confidence level. 
%
%

From Fig.~\ref{fig:one-param-t23delta}, we observe that 
all five one-parameter models could be excluded at $5\sigma$ by T2HK, 
whereas ESSnuSB and DUNE can only exclude Models~1.3 and 1.5 
at $3\sigma$ if the true values of $\sin^2\theta_{23}$ and $\delta_{\rm CP}$ 
are in the vicinity of the current best fit for NO. 
Due to the excellent $\delta_{\rm CP}$ precision of ESSnuSB, 
it is a general trend that it can reject the models in a broader range 
of the $\delta_{\rm CP}$ (true) parameter space than T2HK and DUNE, 
whereas T2HK and DUNE shrink the corresponding intervals 
for the true values of $\theta_{23}$. 
To understand how much the $\theta_{23}$ sensitivity could improve 
if the atmospheric data sample at the far detector is added 
to the ESSnuSB beam data, we add an extra $\chi^2$ for $\sin^2\theta_{23}$ 
with 3~\% precision~\cite{Blennow:2019bvl} to the analysis 
as a placeholder for these data (denoted ``atm proxy''). 
This shows a significant improvement in the sensitivity 
and now ESSnuSB could also exclude Models~1.3 and 1.5 
at approximately $5\sigma$ if the true parameter values 
are the current best-fit values. 
Naturally, the combination of all LBL experiments 
is more sensitive than each individual setup. 
Although Models~1.2 and 1.4 predict $\theta_{23}$ in the higher octant, 
there are also regions in the lower octant, 
where these models cannot be excluded by T2HK and DUNE 
due to the octant degeneracy~\cite{Barger:2001yr,Ghosh:2015ena}. 
For ``ESSnuSB + atm proxy'', there are no degenerate solutions 
for Models~1.2 and 1.4. The reason is that  we consider 
the atmospheric placeholder as an experiment that can only measure 
$\sin^2\theta_{23}$, but for actual atmospheric neutrinos, 
the sensitivity depends on $\sin^2\theta_{23}$ in the appearance channel 
and $\sin^2\theta_{23}$, $\sin^22\theta_{23}$, and $\sin^4\theta_{23}$ 
in the disappearance channel~\cite{Akhmedov:2004ny,Gandhi:2007td,Choubey:2003yp}. 
For LBL combined, the degeneracy is resolved and all models 
could be excluded in most of the parameter space at more than $5\sigma$.  

From Fig.~\ref{fig:two-param-t23delta}, we see that the two-parameter models 
could be excluded at a lower confidence than the one-parameter models 
if the current best-fit parameter values are the true values. 
First, all setups under consideration (including ``ESSnuSB + atm proxy'') 
can exclude Model~2.1 at $3\sigma$ for the current best-fit values.
Second, ESSnuSB and DUNE can exclude Model~2.2 at $3\sigma$, 
whereas T2HK cannot. LBL combined is able to exclude both models 
at $5\sigma$. Finally, none of Models~2.3--2.5 can be excluded at $3\sigma$, 
not even with LBL combined, if the true values of $\sin^2\theta_{23}$ 
and $\delta_\mathrm{CP}$ coincide with the current best fit.
Note that for Model~2.1, which predicts a very narrow range 
of $\sin^2\theta_{23}$ around 0.55, there is a degenerate range around 0.45, which is not resolved even by LBL combined. 
The reason is that this value of $\sin^2\theta_{23}$ is rather close 
to maximal mixing, where the octant sensitivity is generally poor. 
Furthermore, Models~2.4 and 2.5 predict similar values 
of $\theta_{23}$ and $\delta_{\rm CP}$, 
and they therefore give very similar results for the LBL experiments.

\begin{figure*}[t]
\centering
\includegraphics[width=0.45\textwidth]{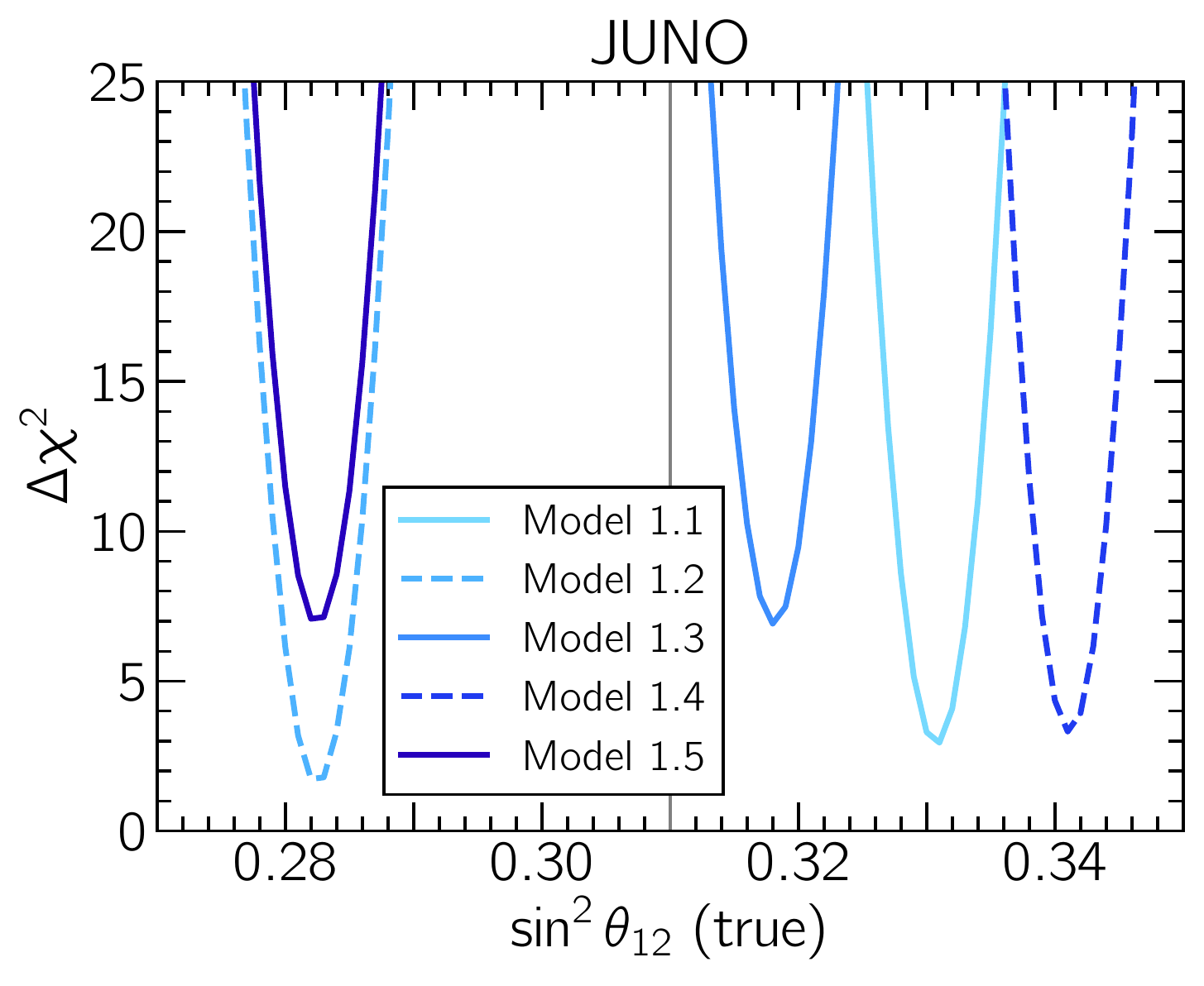}
\hspace{0.05\textwidth}
\includegraphics[width=0.45\textwidth]{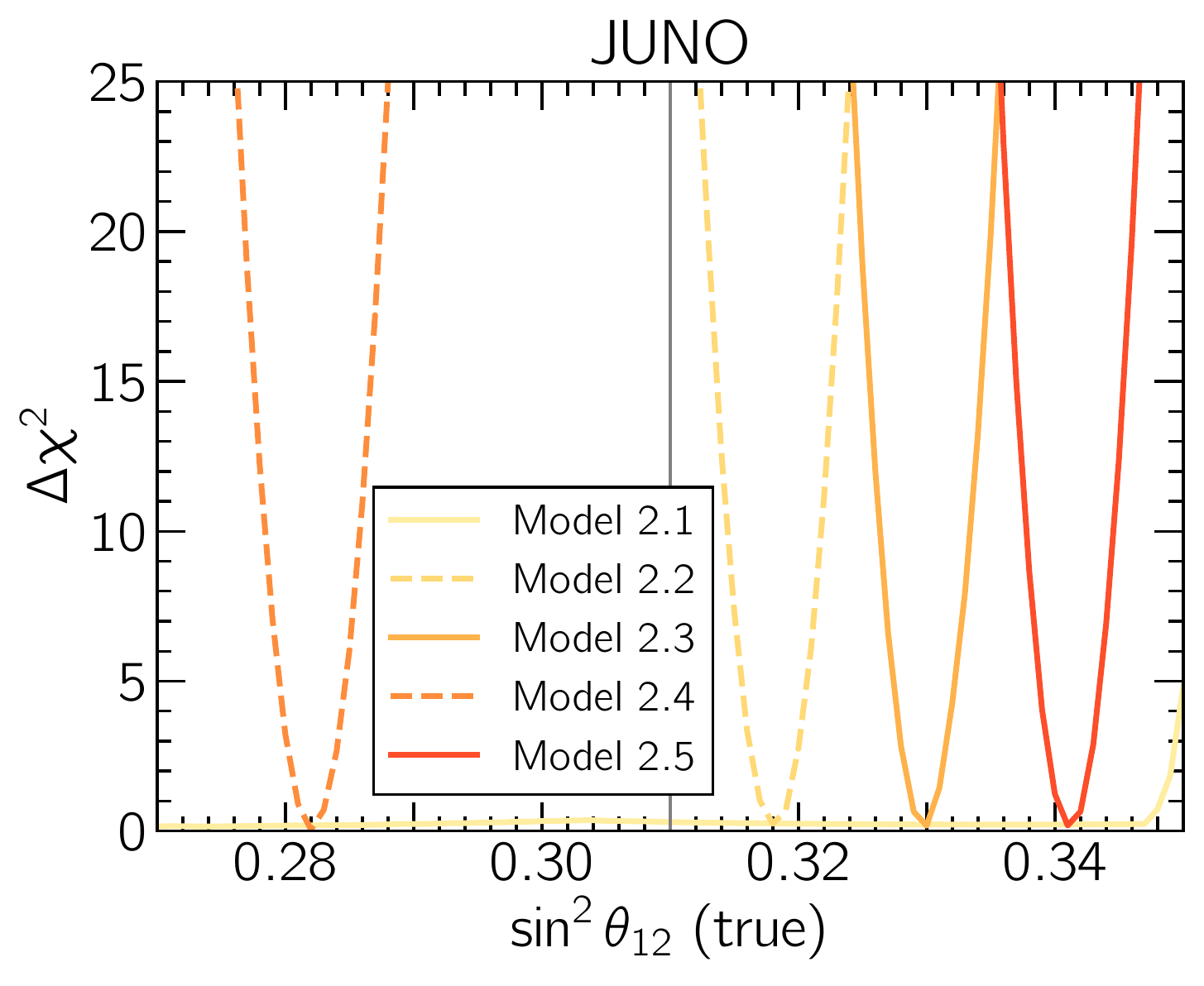}
\caption{Compatibility of one- and two-parameter models with any potentially true value of $\sin^2\theta_{12}$ in the context of JUNO. The \textit{vertical gray line} indicates the present best-fit value of $\sin^2\theta_{12}$ from global neutrino oscillation data.}
\label{fig:JUNO}
\end{figure*}

For JUNO, results in the $\sin^2 \theta_{23}$--$\delta_{\rm CP}$ plane 
would be highly dependent on the true value of $\theta_{12}$. 
The reason is that the expected precision of JUNO for $\sin^2\theta_{12}$ 
is 0.54~\%~\cite{An:2015jdp}, and for the true value 
of $\sin^2 \theta_{12} = 0.310$, all models, except Model~2.1, 
would be excluded at $\Delta\chi^2 > 25$, as estimated from 
$$
\left[\frac{\sin^2 \theta_{12}(\rm model) - 0.310}{0.310 \times 0.0054}\right]^2.
$$ 
This is due to Models~2.2--2.5 predicting $\sin^2 \theta_{12}$ 
in a very narrow range away from the current best-fit value, 
but still within current limits. This is clearer from Fig.~\ref{fig:JUNO},
where we show $\Delta \chi^2$ as a function of the true value 
of $\sin^2\theta_{12}$ for the ten models in the context of JUNO.
Since all models except Model 2.1 predict $\sin^2 \theta_{12}$ 
in very narrow ranges, JUNO is capable of separating the models 
from each other due to its excellent precision on $\sin^2\theta_{12}$. 
For example, Models~2.4 and 2.5, which cannot be separated by LBL combined, 
could be distinguished by JUNO. However, the models that predict similar values 
of $\sin^2 \theta_{12}$ cannot be distinguished by JUNO, 
cf.~Models~1.2, 1.5, and 2.4. 
The flatness of $\Delta \chi^2$ for Model~2.1 is due to not 
making a sharp prediction for $\theta_{12}$. 
As mentioned, if the true value of $\sin^2 \theta_{12}$ is close 
to the current best-fit value, then JUNO will exclude all models 
except Model~2.1 at more than $\Delta\chi^2 = 25$. 
Note that for the two-parameter models the minimal values of $\Delta\chi^2$ 
are close to zero, whereas the minimum $\Delta \chi^2$ is 
2.8, 1.6, 6.9, 3.2, and 6.9 for Models~1.1--1.5, respectively.  
These values are due to the $\theta_{23}$ pull given by 
$$
\left[\frac{\sin^2\theta_{23}(\rm model)-0.563}{0.024}\right]^2,
$$ 
which is close to zero for the two-parameter models.
%

\section{Summary and Conclusions}
\label{sec:conclusions}
%
We have explored the potential of future neutrino experiments---%
ESSnuSB, T2HK, DUNE, and JUNO---to test lepton flavor models 
based on non-Abelian discrete symmetries. 
Such models lead either to sharp predictions for 
or correlations among the leptonic mixing parameters.
The results obtained show that the high-precision measurement 
of $\sin^2\theta_{12}$ by JUNO will be crucial in discriminating among 
and excluding most of the considered models. 
For instance, if the true value of $\sin^2\theta_{12}$ is slightly larger 
than its present best-fit value (say by $\sim0.01$), 
only Models 1.3, 2.1, and 2.2 would survive at $3\sigma$. 
In addition, if the true value of $\sin^2\theta_{23}$ occurs in the vicinity 
of its current best-fit value, Model 1.3 will be excluded at more than $3\sigma$ 
by the LBL setups, whereas the viability of Models 2.1 and 2.2 
will depend on the true value of $\delta_\mathrm{CP}$. 
Our investigation shows that all ten models will be excluded 
at more than $5 \sigma$ (Model~2.1 by LBL combined 
and the other models by JUNO) if the true values 
of the leptonic mixing parameters occur close to their present best-fit values.
In conclusion, our results demonstrate that the complementarity 
between accelerator and reactor experiments provides 
a unique and extremely powerful utility for pinning down 
this broad class of lepton flavor models.

\begin{acknowledgments}
We would like to thank Patrick Huber and Rebekah Pestes for supplying us with their~.glb file for the JUNO experiment. We would also like to thank Marcos Dracos, Tord Ekel{\"o}f, Enrique Fernandez-Martinez, and Suprabh Prakash for useful discussions and Marie-Laure Schneider for comments on our work. This project is supported by the COST Action CA15139 {\it ``Combining forces for a novel European facility for neutrino-antineutrino symmetry-violation discovery''} (EuroNuNet). It has also received funding from the European Union's Horizon 2020 research and innovation programme under grant agreement No.~777419. T.O.~acknowledges support by the Swedish Research Council (Vetenskaps\-r{\aa}det) through Contract No.~2017-03934 and the KTH Royal Institute of Technology for a sabbatical period at the University of Iceland.
\end{acknowledgments}

\bibliography{Flavor_Models_at_Neutrino_Experiments_v2}

\begin{thebibliography}{61}%
\makeatletter
\providecommand \@ifxundefined [1]{%
 \@ifx{#1\undefined}
}%
\providecommand \@ifnum [1]{%
 \ifnum #1\expandafter \@firstoftwo
 \else \expandafter \@secondoftwo
 \fi
}%
\providecommand \@ifx [1]{%
 \ifx #1\expandafter \@firstoftwo
 \else \expandafter \@secondoftwo
 \fi
}%
\providecommand \natexlab [1]{#1}%
\providecommand \enquote  [1]{``#1''}%
\providecommand \bibnamefont  [1]{#1}%
\providecommand \bibfnamefont [1]{#1}%
\providecommand \citenamefont [1]{#1}%
\providecommand \href@noop [0]{\@secondoftwo}%
\providecommand \href [0]{\begingroup \@sanitize@url \@href}%
\providecommand \@href[1]{\@@startlink{#1}\@@href}%
\providecommand \@@href[1]{\endgroup#1\@@endlink}%
\providecommand \@sanitize@url [0]{\catcode `\\12\catcode `\$12\catcode
  `\&12\catcode `\#12\catcode `\^12\catcode `\_12\catcode `\%12\relax}%
\providecommand \@@startlink[1]{}%
\providecommand \@@endlink[0]{}%
\providecommand \url  [0]{\begingroup\@sanitize@url \@url }%
\providecommand \@url [1]{\endgroup\@href {#1}{\urlprefix }}%
\providecommand \urlprefix  [0]{URL }%
\providecommand \Eprint [0]{\href }%
\providecommand \doibase [0]{http://dx.doi.org/}%
\providecommand \selectlanguage [0]{\@gobble}%
\providecommand \bibinfo  [0]{\@secondoftwo}%
\providecommand \bibfield  [0]{\@secondoftwo}%
\providecommand \translation [1]{[#1]}%
\providecommand \BibitemOpen [0]{}%
\providecommand \bibitemStop [0]{}%
\providecommand \bibitemNoStop [0]{.\EOS\space}%
\providecommand \EOS [0]{\spacefactor3000\relax}%
\providecommand \BibitemShut  [1]{\csname bibitem#1\endcsname}%
\let\auto@bib@innerbib\@empty
\bibitem [{\citenamefont {Altarelli}\ and\ \citenamefont
  {Feruglio}(2010)}]{Altarelli:2010gt}%
  \BibitemOpen
  \bibfield  {author} {\bibinfo {author} {\bibfnamefont {G.}~\bibnamefont
  {Altarelli}}\ and\ \bibinfo {author} {\bibfnamefont {F.}~\bibnamefont
  {Feruglio}},\ }\href {\doibase 10.1103/RevModPhys.82.2701} {\bibfield
  {journal} {\bibinfo  {journal} {Rev. Mod. Phys.}\ }\textbf {\bibinfo {volume}
  {82}},\ \bibinfo {pages} {2701} (\bibinfo {year} {2010})},\ \Eprint
  {http://arxiv.org/abs/1002.0211} {arXiv:1002.0211 [hep-ph]} \BibitemShut
  {NoStop}%
\bibitem [{\citenamefont {Ishimori}\ \emph {et~al.}(2010)\citenamefont
  {Ishimori}, \citenamefont {Kobayashi}, \citenamefont {Ohki}, \citenamefont
  {Shimizu}, \citenamefont {Okada},\ and\ \citenamefont
  {Tanimoto}}]{Ishimori:2010au}%
  \BibitemOpen
  \bibfield  {author} {\bibinfo {author} {\bibfnamefont {H.}~\bibnamefont
  {Ishimori}}, \bibinfo {author} {\bibfnamefont {T.}~\bibnamefont {Kobayashi}},
  \bibinfo {author} {\bibfnamefont {H.}~\bibnamefont {Ohki}}, \bibinfo {author}
  {\bibfnamefont {Y.}~\bibnamefont {Shimizu}}, \bibinfo {author} {\bibfnamefont
  {H.}~\bibnamefont {Okada}}, \ and\ \bibinfo {author} {\bibfnamefont
  {M.}~\bibnamefont {Tanimoto}},\ }\href {\doibase 10.1143/PTPS.183.1}
  {\bibfield  {journal} {\bibinfo  {journal} {Prog. Theor. Phys. Suppl.}\
  }\textbf {\bibinfo {volume} {183}},\ \bibinfo {pages} {1} (\bibinfo {year}
  {2010})},\ \Eprint {http://arxiv.org/abs/1003.3552} {arXiv:1003.3552
  [hep-th]} \BibitemShut {NoStop}%
\bibitem [{\citenamefont {King}\ and\ \citenamefont
  {Luhn}(2013)}]{King:2013eh}%
  \BibitemOpen
  \bibfield  {author} {\bibinfo {author} {\bibfnamefont {S.~F.}\ \bibnamefont
  {King}}\ and\ \bibinfo {author} {\bibfnamefont {C.}~\bibnamefont {Luhn}},\
  }\href {\doibase 10.1088/0034-4885/76/5/056201} {\bibfield  {journal}
  {\bibinfo  {journal} {Rept. Prog. Phys.}\ }\textbf {\bibinfo {volume} {76}},\
  \bibinfo {pages} {056201} (\bibinfo {year} {2013})},\ \Eprint
  {http://arxiv.org/abs/1301.1340} {arXiv:1301.1340 [hep-ph]} \BibitemShut
  {NoStop}%
\bibitem [{\citenamefont {Petcov}(2018)}]{Petcov:2017ggy}%
  \BibitemOpen
  \bibfield  {author} {\bibinfo {author} {\bibfnamefont {S.~T.}\ \bibnamefont
  {Petcov}},\ }\href {\doibase 10.1140/epjc/s10052-018-6158-5} {\bibfield
  {journal} {\bibinfo  {journal} {Eur. Phys. J.}\ }\textbf {\bibinfo {volume}
  {C78}},\ \bibinfo {pages} {709} (\bibinfo {year} {2018})},\ \Eprint
  {http://arxiv.org/abs/1711.10806} {arXiv:1711.10806 [hep-ph]} \BibitemShut
  {NoStop}%
\bibitem [{\citenamefont {Feruglio}\ and\ \citenamefont
  {Romanino}(2019)}]{Feruglio:2019ktm}%
  \BibitemOpen
  \bibfield  {author} {\bibinfo {author} {\bibfnamefont {F.}~\bibnamefont
  {Feruglio}}\ and\ \bibinfo {author} {\bibfnamefont {A.}~\bibnamefont
  {Romanino}},\ }\href@noop {} {\  (\bibinfo {year} {2019})},\ \Eprint
  {http://arxiv.org/abs/1912.06028} {arXiv:1912.06028 [hep-ph]} \BibitemShut
  {NoStop}%
\bibitem [{\citenamefont {Abe}\ \emph {et~al.}(2020)\citenamefont {Abe} \emph
  {et~al.}}]{Abe:2019vii}%
  \BibitemOpen
  \bibfield  {author} {\bibinfo {author} {\bibfnamefont {K.}~\bibnamefont
  {Abe}} \emph {et~al.} (\bibinfo {collaboration} {T2K}),\ }\href {\doibase
  10.1038/s41586-020-2177-0} {\bibfield  {journal} {\bibinfo  {journal}
  {Nature}\ }\textbf {\bibinfo {volume} {580}},\ \bibinfo {pages} {339}
  (\bibinfo {year} {2020})},\ \bibinfo {note} {[Erratum: Nature 583, E16
  (2020)]},\ \Eprint {http://arxiv.org/abs/1910.03887} {arXiv:1910.03887
  [hep-ex]} \BibitemShut {NoStop}%
\bibitem [{\citenamefont {Acero}\ \emph {et~al.}(2019)\citenamefont {Acero}
  \emph {et~al.}}]{Acero:2019ksn}%
  \BibitemOpen
  \bibfield  {author} {\bibinfo {author} {\bibfnamefont {M.~A.}\ \bibnamefont
  {Acero}} \emph {et~al.} (\bibinfo {collaboration} {NOvA}),\ }\href {\doibase
  10.1103/PhysRevLett.123.151803} {\bibfield  {journal} {\bibinfo  {journal}
  {Phys. Rev. Lett.}\ }\textbf {\bibinfo {volume} {123}},\ \bibinfo {pages}
  {151803} (\bibinfo {year} {2019})},\ \Eprint
  {http://arxiv.org/abs/1906.04907} {arXiv:1906.04907 [hep-ex]} \BibitemShut
  {NoStop}%
\bibitem [{\citenamefont {Antusch}\ \emph {et~al.}(2007)\citenamefont
  {Antusch}, \citenamefont {Huber}, \citenamefont {King},\ and\ \citenamefont
  {Schwetz}}]{Antusch:2007rk}%
  \BibitemOpen
  \bibfield  {author} {\bibinfo {author} {\bibfnamefont {S.}~\bibnamefont
  {Antusch}}, \bibinfo {author} {\bibfnamefont {P.}~\bibnamefont {Huber}},
  \bibinfo {author} {\bibfnamefont {S.~F.}\ \bibnamefont {King}}, \ and\
  \bibinfo {author} {\bibfnamefont {T.}~\bibnamefont {Schwetz}},\ }\href
  {\doibase 10.1088/1126-6708/2007/04/060} {\bibfield  {journal} {\bibinfo
  {journal} {JHEP}\ }\textbf {\bibinfo {volume} {04}},\ \bibinfo {pages} {060}
  (\bibinfo {year} {2007})},\ \Eprint {http://arxiv.org/abs/hep-ph/0702286}
  {arXiv:hep-ph/0702286 [HEP-PH]} \BibitemShut {NoStop}%
\bibitem [{\citenamefont {Ballett}\ \emph
  {et~al.}(2014{\natexlab{a}})\citenamefont {Ballett}, \citenamefont {King},
  \citenamefont {Luhn}, \citenamefont {Pascoli},\ and\ \citenamefont
  {Schmidt}}]{Ballett:2013wya}%
  \BibitemOpen
  \bibfield  {author} {\bibinfo {author} {\bibfnamefont {P.}~\bibnamefont
  {Ballett}}, \bibinfo {author} {\bibfnamefont {S.~F.}\ \bibnamefont {King}},
  \bibinfo {author} {\bibfnamefont {C.}~\bibnamefont {Luhn}}, \bibinfo {author}
  {\bibfnamefont {S.}~\bibnamefont {Pascoli}}, \ and\ \bibinfo {author}
  {\bibfnamefont {M.~A.}\ \bibnamefont {Schmidt}},\ }\href {\doibase
  10.1103/PhysRevD.89.016016} {\bibfield  {journal} {\bibinfo  {journal} {Phys.
  Rev.}\ }\textbf {\bibinfo {volume} {D89}},\ \bibinfo {pages} {016016}
  (\bibinfo {year} {2014}{\natexlab{a}})},\ \Eprint
  {http://arxiv.org/abs/1308.4314} {arXiv:1308.4314 [hep-ph]} \BibitemShut
  {NoStop}%
\bibitem [{\citenamefont {Petcov}(2015)}]{Petcov:2014laa}%
  \BibitemOpen
  \bibfield  {author} {\bibinfo {author} {\bibfnamefont {S.~T.}\ \bibnamefont
  {Petcov}},\ }\href {\doibase 10.1016/j.nuclphysb.2015.01.011} {\bibfield
  {journal} {\bibinfo  {journal} {Nucl. Phys.}\ }\textbf {\bibinfo {volume}
  {B892}},\ \bibinfo {pages} {400} (\bibinfo {year} {2015})},\ \Eprint
  {http://arxiv.org/abs/1405.6006} {arXiv:1405.6006 [hep-ph]} \BibitemShut
  {NoStop}%
\bibitem [{\citenamefont {Girardi}\ \emph
  {et~al.}(2015{\natexlab{a}})\citenamefont {Girardi}, \citenamefont {Petcov},\
  and\ \citenamefont {Titov}}]{Girardi:2014faa}%
  \BibitemOpen
  \bibfield  {author} {\bibinfo {author} {\bibfnamefont {I.}~\bibnamefont
  {Girardi}}, \bibinfo {author} {\bibfnamefont {S.~T.}\ \bibnamefont {Petcov}},
  \ and\ \bibinfo {author} {\bibfnamefont {A.~V.}\ \bibnamefont {Titov}},\
  }\href {\doibase 10.1016/j.nuclphysb.2015.03.026} {\bibfield  {journal}
  {\bibinfo  {journal} {Nucl. Phys.}\ }\textbf {\bibinfo {volume} {B894}},\
  \bibinfo {pages} {733} (\bibinfo {year} {2015}{\natexlab{a}})},\ \Eprint
  {http://arxiv.org/abs/1410.8056} {arXiv:1410.8056 [hep-ph]} \BibitemShut
  {NoStop}%
\bibitem [{\citenamefont {Ballett}\ \emph
  {et~al.}(2014{\natexlab{b}})\citenamefont {Ballett}, \citenamefont {King},
  \citenamefont {Luhn}, \citenamefont {Pascoli},\ and\ \citenamefont
  {Schmidt}}]{Ballett:2014dua}%
  \BibitemOpen
  \bibfield  {author} {\bibinfo {author} {\bibfnamefont {P.}~\bibnamefont
  {Ballett}}, \bibinfo {author} {\bibfnamefont {S.~F.}\ \bibnamefont {King}},
  \bibinfo {author} {\bibfnamefont {C.}~\bibnamefont {Luhn}}, \bibinfo {author}
  {\bibfnamefont {S.}~\bibnamefont {Pascoli}}, \ and\ \bibinfo {author}
  {\bibfnamefont {M.~A.}\ \bibnamefont {Schmidt}},\ }\href {\doibase
  10.1007/JHEP12(2014)122} {\bibfield  {journal} {\bibinfo  {journal} {JHEP}\
  }\textbf {\bibinfo {volume} {12}},\ \bibinfo {pages} {122} (\bibinfo {year}
  {2014}{\natexlab{b}})},\ \Eprint {http://arxiv.org/abs/1410.7573}
  {arXiv:1410.7573 [hep-ph]} \BibitemShut {NoStop}%
\bibitem [{\citenamefont {Girardi}\ \emph
  {et~al.}(2015{\natexlab{b}})\citenamefont {Girardi}, \citenamefont {Petcov},\
  and\ \citenamefont {Titov}}]{Girardi:2015vha}%
  \BibitemOpen
  \bibfield  {author} {\bibinfo {author} {\bibfnamefont {I.}~\bibnamefont
  {Girardi}}, \bibinfo {author} {\bibfnamefont {S.~T.}\ \bibnamefont {Petcov}},
  \ and\ \bibinfo {author} {\bibfnamefont {A.~V.}\ \bibnamefont {Titov}},\
  }\href {\doibase 10.1140/epjc/s10052-015-3559-6} {\bibfield  {journal}
  {\bibinfo  {journal} {Eur. Phys. J.}\ }\textbf {\bibinfo {volume} {C75}},\
  \bibinfo {pages} {345} (\bibinfo {year} {2015}{\natexlab{b}})},\ \Eprint
  {http://arxiv.org/abs/1504.00658} {arXiv:1504.00658 [hep-ph]} \BibitemShut
  {NoStop}%
\bibitem [{\citenamefont {Agarwalla}\ \emph {et~al.}(2018)\citenamefont
  {Agarwalla}, \citenamefont {Chatterjee}, \citenamefont {Petcov},\ and\
  \citenamefont {Titov}}]{Agarwalla:2017wct}%
  \BibitemOpen
  \bibfield  {author} {\bibinfo {author} {\bibfnamefont {S.~K.}\ \bibnamefont
  {Agarwalla}}, \bibinfo {author} {\bibfnamefont {S.~S.}\ \bibnamefont
  {Chatterjee}}, \bibinfo {author} {\bibfnamefont {S.~T.}\ \bibnamefont
  {Petcov}}, \ and\ \bibinfo {author} {\bibfnamefont {A.~V.}\ \bibnamefont
  {Titov}},\ }\href {\doibase 10.1140/epjc/s10052-018-5772-6} {\bibfield
  {journal} {\bibinfo  {journal} {Eur. Phys. J.}\ }\textbf {\bibinfo {volume}
  {C78}},\ \bibinfo {pages} {286} (\bibinfo {year} {2018})},\ \Eprint
  {http://arxiv.org/abs/1711.02107} {arXiv:1711.02107 [hep-ph]} \BibitemShut
  {NoStop}%
\bibitem [{\citenamefont {Petcov}\ and\ \citenamefont
  {Titov}(2018)}]{Petcov:2018snn}%
  \BibitemOpen
  \bibfield  {author} {\bibinfo {author} {\bibfnamefont {S.~T.}\ \bibnamefont
  {Petcov}}\ and\ \bibinfo {author} {\bibfnamefont {A.~V.}\ \bibnamefont
  {Titov}},\ }\href {\doibase 10.1103/PhysRevD.97.115045} {\bibfield  {journal}
  {\bibinfo  {journal} {Phys. Rev.}\ }\textbf {\bibinfo {volume} {D97}},\
  \bibinfo {pages} {115045} (\bibinfo {year} {2018})},\ \Eprint
  {http://arxiv.org/abs/1804.00182} {arXiv:1804.00182 [hep-ph]} \BibitemShut
  {NoStop}%
\bibitem [{\citenamefont {Blennow}\ \emph
  {et~al.}(2020{\natexlab{a}})\citenamefont {Blennow}, \citenamefont {Ghosh},
  \citenamefont {Ohlsson},\ and\ \citenamefont {Titov}}]{Blennow:2020snb}%
  \BibitemOpen
  \bibfield  {author} {\bibinfo {author} {\bibfnamefont {M.}~\bibnamefont
  {Blennow}}, \bibinfo {author} {\bibfnamefont {M.}~\bibnamefont {Ghosh}},
  \bibinfo {author} {\bibfnamefont {T.}~\bibnamefont {Ohlsson}}, \ and\
  \bibinfo {author} {\bibfnamefont {A.}~\bibnamefont {Titov}},\ }\href
  {\doibase 10.1007/JHEP07(2020)014} {\bibfield  {journal} {\bibinfo  {journal}
  {JHEP}\ }\textbf {\bibinfo {volume} {07}},\ \bibinfo {pages} {014} (\bibinfo
  {year} {2020}{\natexlab{a}})},\ \Eprint {http://arxiv.org/abs/2004.00017}
  {arXiv:2004.00017 [hep-ph]} \BibitemShut {NoStop}%
\bibitem [{\citenamefont {Babu}\ \emph {et~al.}(2003)\citenamefont {Babu},
  \citenamefont {Ma},\ and\ \citenamefont {Valle}}]{Babu:2002dz}%
  \BibitemOpen
  \bibfield  {author} {\bibinfo {author} {\bibfnamefont {K.}~\bibnamefont
  {Babu}}, \bibinfo {author} {\bibfnamefont {E.}~\bibnamefont {Ma}}, \ and\
  \bibinfo {author} {\bibfnamefont {J.}~\bibnamefont {Valle}},\ }\href
  {\doibase 10.1016/S0370-2693(02)03153-2} {\bibfield  {journal} {\bibinfo
  {journal} {Phys. Lett. B}\ }\textbf {\bibinfo {volume} {552}},\ \bibinfo
  {pages} {207} (\bibinfo {year} {2003})},\ \Eprint
  {http://arxiv.org/abs/hep-ph/0206292} {arXiv:hep-ph/0206292} \BibitemShut
  {NoStop}%
\bibitem [{\citenamefont {He}\ \emph {et~al.}(2006)\citenamefont {He},
  \citenamefont {Keum},\ and\ \citenamefont {Volkas}}]{He:2006dk}%
  \BibitemOpen
  \bibfield  {author} {\bibinfo {author} {\bibfnamefont {X.-G.}\ \bibnamefont
  {He}}, \bibinfo {author} {\bibfnamefont {Y.-Y.}\ \bibnamefont {Keum}}, \ and\
  \bibinfo {author} {\bibfnamefont {R.~R.}\ \bibnamefont {Volkas}},\ }\href
  {\doibase 10.1088/1126-6708/2006/04/039} {\bibfield  {journal} {\bibinfo
  {journal} {JHEP}\ }\textbf {\bibinfo {volume} {04}},\ \bibinfo {pages} {039}
  (\bibinfo {year} {2006})},\ \Eprint {http://arxiv.org/abs/hep-ph/0601001}
  {arXiv:hep-ph/0601001} \BibitemShut {NoStop}%
\bibitem [{\citenamefont {Ma}\ \emph {et~al.}(2006)\citenamefont {Ma},
  \citenamefont {Sawanaka},\ and\ \citenamefont {Tanimoto}}]{Ma:2006sk}%
  \BibitemOpen
  \bibfield  {author} {\bibinfo {author} {\bibfnamefont {E.}~\bibnamefont
  {Ma}}, \bibinfo {author} {\bibfnamefont {H.}~\bibnamefont {Sawanaka}}, \ and\
  \bibinfo {author} {\bibfnamefont {M.}~\bibnamefont {Tanimoto}},\ }\href
  {\doibase 10.1016/j.physletb.2006.08.062} {\bibfield  {journal} {\bibinfo
  {journal} {Phys. Lett. B}\ }\textbf {\bibinfo {volume} {641}},\ \bibinfo
  {pages} {301} (\bibinfo {year} {2006})},\ \Eprint
  {http://arxiv.org/abs/hep-ph/0606103} {arXiv:hep-ph/0606103} \BibitemShut
  {NoStop}%
\bibitem [{\citenamefont {King}\ and\ \citenamefont
  {Malinsky}(2007)}]{King:2006np}%
  \BibitemOpen
  \bibfield  {author} {\bibinfo {author} {\bibfnamefont {S.~F.}\ \bibnamefont
  {King}}\ and\ \bibinfo {author} {\bibfnamefont {M.}~\bibnamefont
  {Malinsky}},\ }\href {\doibase 10.1016/j.physletb.2006.12.006} {\bibfield
  {journal} {\bibinfo  {journal} {Phys. Lett. B}\ }\textbf {\bibinfo {volume}
  {645}},\ \bibinfo {pages} {351} (\bibinfo {year} {2007})},\ \Eprint
  {http://arxiv.org/abs/hep-ph/0610250} {arXiv:hep-ph/0610250} \BibitemShut
  {NoStop}%
\bibitem [{\citenamefont {Feruglio}\ \emph {et~al.}(2007)\citenamefont
  {Feruglio}, \citenamefont {Hagedorn}, \citenamefont {Lin},\ and\
  \citenamefont {Merlo}}]{Feruglio:2007uu}%
  \BibitemOpen
  \bibfield  {author} {\bibinfo {author} {\bibfnamefont {F.}~\bibnamefont
  {Feruglio}}, \bibinfo {author} {\bibfnamefont {C.}~\bibnamefont {Hagedorn}},
  \bibinfo {author} {\bibfnamefont {Y.}~\bibnamefont {Lin}}, \ and\ \bibinfo
  {author} {\bibfnamefont {L.}~\bibnamefont {Merlo}},\ }\href {\doibase
  10.1016/j.nuclphysb.2007.04.002} {\bibfield  {journal} {\bibinfo  {journal}
  {Nucl. Phys. B}\ }\textbf {\bibinfo {volume} {775}},\ \bibinfo {pages} {120}
  (\bibinfo {year} {2007})},\ \bibinfo {note} {[Erratum: Nucl.Phys.B 836,
  127--128 (2010)]},\ \Eprint {http://arxiv.org/abs/hep-ph/0702194}
  {arXiv:hep-ph/0702194} \BibitemShut {NoStop}%
\bibitem [{\citenamefont {Chen}\ and\ \citenamefont
  {Mahanthappa}(2007)}]{Chen:2007afa}%
  \BibitemOpen
  \bibfield  {author} {\bibinfo {author} {\bibfnamefont {M.-C.}\ \bibnamefont
  {Chen}}\ and\ \bibinfo {author} {\bibfnamefont {K.}~\bibnamefont
  {Mahanthappa}},\ }\href {\doibase 10.1016/j.physletb.2007.06.064} {\bibfield
  {journal} {\bibinfo  {journal} {Phys. Lett. B}\ }\textbf {\bibinfo {volume}
  {652}},\ \bibinfo {pages} {34} (\bibinfo {year} {2007})},\ \Eprint
  {http://arxiv.org/abs/0705.0714} {arXiv:0705.0714 [hep-ph]} \BibitemShut
  {NoStop}%
\bibitem [{\citenamefont {Frampton}\ and\ \citenamefont
  {Kephart}(2007)}]{Frampton:2007et}%
  \BibitemOpen
  \bibfield  {author} {\bibinfo {author} {\bibfnamefont {P.~H.}\ \bibnamefont
  {Frampton}}\ and\ \bibinfo {author} {\bibfnamefont {T.~W.}\ \bibnamefont
  {Kephart}},\ }\href {\doibase 10.1088/1126-6708/2007/09/110} {\bibfield
  {journal} {\bibinfo  {journal} {JHEP}\ }\textbf {\bibinfo {volume} {09}},\
  \bibinfo {pages} {110} (\bibinfo {year} {2007})},\ \Eprint
  {http://arxiv.org/abs/0706.1186} {arXiv:0706.1186 [hep-ph]} \BibitemShut
  {NoStop}%
\bibitem [{\citenamefont {Feruglio}\ \emph {et~al.}(2013)\citenamefont
  {Feruglio}, \citenamefont {Hagedorn},\ and\ \citenamefont
  {Ziegler}}]{Feruglio:2012cw}%
  \BibitemOpen
  \bibfield  {author} {\bibinfo {author} {\bibfnamefont {F.}~\bibnamefont
  {Feruglio}}, \bibinfo {author} {\bibfnamefont {C.}~\bibnamefont {Hagedorn}},
  \ and\ \bibinfo {author} {\bibfnamefont {R.}~\bibnamefont {Ziegler}},\ }\href
  {\doibase 10.1007/JHEP07(2013)027} {\bibfield  {journal} {\bibinfo  {journal}
  {JHEP}\ }\textbf {\bibinfo {volume} {07}},\ \bibinfo {pages} {027} (\bibinfo
  {year} {2013})},\ \Eprint {http://arxiv.org/abs/1211.5560} {arXiv:1211.5560
  [hep-ph]} \BibitemShut {NoStop}%
\bibitem [{\citenamefont {Li}\ and\ \citenamefont {Ding}(2015)}]{Li:2015jxa}%
  \BibitemOpen
  \bibfield  {author} {\bibinfo {author} {\bibfnamefont {C.-C.}\ \bibnamefont
  {Li}}\ and\ \bibinfo {author} {\bibfnamefont {G.-J.}\ \bibnamefont {Ding}},\
  }\href {\doibase 10.1007/JHEP05(2015)100} {\bibfield  {journal} {\bibinfo
  {journal} {JHEP}\ }\textbf {\bibinfo {volume} {05}},\ \bibinfo {pages} {100}
  (\bibinfo {year} {2015})},\ \Eprint {http://arxiv.org/abs/1503.03711}
  {arXiv:1503.03711 [hep-ph]} \BibitemShut {NoStop}%
\bibitem [{\citenamefont {Di~Iura}\ \emph {et~al.}(2015)\citenamefont
  {Di~Iura}, \citenamefont {Hagedorn},\ and\ \citenamefont
  {Meloni}}]{DiIura:2015kfa}%
  \BibitemOpen
  \bibfield  {author} {\bibinfo {author} {\bibfnamefont {A.}~\bibnamefont
  {Di~Iura}}, \bibinfo {author} {\bibfnamefont {C.}~\bibnamefont {Hagedorn}}, \
  and\ \bibinfo {author} {\bibfnamefont {D.}~\bibnamefont {Meloni}},\ }\href
  {\doibase 10.1007/JHEP08(2015)037} {\bibfield  {journal} {\bibinfo  {journal}
  {JHEP}\ }\textbf {\bibinfo {volume} {08}},\ \bibinfo {pages} {037} (\bibinfo
  {year} {2015})},\ \Eprint {http://arxiv.org/abs/1503.04140} {arXiv:1503.04140
  [hep-ph]} \BibitemShut {NoStop}%
\bibitem [{\citenamefont {Ballett}\ \emph {et~al.}(2015)\citenamefont
  {Ballett}, \citenamefont {Pascoli},\ and\ \citenamefont
  {Turner}}]{Ballett:2015wia}%
  \BibitemOpen
  \bibfield  {author} {\bibinfo {author} {\bibfnamefont {P.}~\bibnamefont
  {Ballett}}, \bibinfo {author} {\bibfnamefont {S.}~\bibnamefont {Pascoli}}, \
  and\ \bibinfo {author} {\bibfnamefont {J.}~\bibnamefont {Turner}},\ }\href
  {\doibase 10.1103/PhysRevD.92.093008} {\bibfield  {journal} {\bibinfo
  {journal} {Phys. Rev.}\ }\textbf {\bibinfo {volume} {D92}},\ \bibinfo {pages}
  {093008} (\bibinfo {year} {2015})},\ \Eprint
  {http://arxiv.org/abs/1503.07543} {arXiv:1503.07543 [hep-ph]} \BibitemShut
  {NoStop}%
\bibitem [{\citenamefont {Girardi}\ \emph {et~al.}(2016)\citenamefont
  {Girardi}, \citenamefont {Petcov}, \citenamefont {Stuart},\ and\
  \citenamefont {Titov}}]{Girardi:2015rwa}%
  \BibitemOpen
  \bibfield  {author} {\bibinfo {author} {\bibfnamefont {I.}~\bibnamefont
  {Girardi}}, \bibinfo {author} {\bibfnamefont {S.~T.}\ \bibnamefont {Petcov}},
  \bibinfo {author} {\bibfnamefont {A.~J.}\ \bibnamefont {Stuart}}, \ and\
  \bibinfo {author} {\bibfnamefont {A.~V.}\ \bibnamefont {Titov}},\ }\href
  {\doibase 10.1016/j.nuclphysb.2015.10.020} {\bibfield  {journal} {\bibinfo
  {journal} {Nucl. Phys.}\ }\textbf {\bibinfo {volume} {B902}},\ \bibinfo
  {pages} {1} (\bibinfo {year} {2016})},\ \Eprint
  {http://arxiv.org/abs/1509.02502} {arXiv:1509.02502 [hep-ph]} \BibitemShut
  {NoStop}%
\bibitem [{\citenamefont {Esteban}\ \emph {et~al.}(2019)\citenamefont
  {Esteban}, \citenamefont {Gonzalez-Garcia}, \citenamefont
  {Hernandez-Cabezudo}, \citenamefont {Maltoni},\ and\ \citenamefont
  {Schwetz}}]{Esteban:2018azc}%
  \BibitemOpen
  \bibfield  {author} {\bibinfo {author} {\bibfnamefont {I.}~\bibnamefont
  {Esteban}}, \bibinfo {author} {\bibfnamefont {M.~C.}\ \bibnamefont
  {Gonzalez-Garcia}}, \bibinfo {author} {\bibfnamefont {A.}~\bibnamefont
  {Hernandez-Cabezudo}}, \bibinfo {author} {\bibfnamefont {M.}~\bibnamefont
  {Maltoni}}, \ and\ \bibinfo {author} {\bibfnamefont {T.}~\bibnamefont
  {Schwetz}},\ }\href {\doibase 10.1007/JHEP01(2019)106} {\bibfield  {journal}
  {\bibinfo  {journal} {JHEP}\ }\textbf {\bibinfo {volume} {01}},\ \bibinfo
  {pages} {106} (\bibinfo {year} {2019})},\ \Eprint
  {http://arxiv.org/abs/1811.05487} {arXiv:1811.05487 [hep-ph]} \BibitemShut
  {NoStop}%
\bibitem [{\citenamefont {Esteban}\ \emph {et~al.}()\citenamefont {Esteban},
  \citenamefont {Gonzalez-Garcia}, \citenamefont {Hernandez-Cabezudo},
  \citenamefont {Maltoni},\ and\ \citenamefont {Schwetz}}]{NuFITv41}%
  \BibitemOpen
  \bibfield  {author} {\bibinfo {author} {\bibfnamefont {I.}~\bibnamefont
  {Esteban}}, \bibinfo {author} {\bibfnamefont {M.~C.}\ \bibnamefont
  {Gonzalez-Garcia}}, \bibinfo {author} {\bibfnamefont {A.}~\bibnamefont
  {Hernandez-Cabezudo}}, \bibinfo {author} {\bibfnamefont {M.}~\bibnamefont
  {Maltoni}}, \ and\ \bibinfo {author} {\bibfnamefont {T.}~\bibnamefont
  {Schwetz}},\ }\href@noop {} {\enquote {\bibinfo {title} {{NuFIT 4.1:
  Three-neutrino fit based on data available in July 2019}},}\ }\bibinfo
  {howpublished} {{\href{http://www.nu-fit.org}{www.nu-fit.org}}}\BibitemShut
  {NoStop}%
\bibitem [{\citenamefont {Capozzi}\ \emph {et~al.}(2018)\citenamefont
  {Capozzi}, \citenamefont {Lisi}, \citenamefont {Marrone},\ and\ \citenamefont
  {Palazzo}}]{Capozzi:2018ubv}%
  \BibitemOpen
  \bibfield  {author} {\bibinfo {author} {\bibfnamefont {F.}~\bibnamefont
  {Capozzi}}, \bibinfo {author} {\bibfnamefont {E.}~\bibnamefont {Lisi}},
  \bibinfo {author} {\bibfnamefont {A.}~\bibnamefont {Marrone}}, \ and\
  \bibinfo {author} {\bibfnamefont {A.}~\bibnamefont {Palazzo}},\ }\href
  {\doibase 10.1016/j.ppnp.2018.05.005} {\bibfield  {journal} {\bibinfo
  {journal} {Prog. Part. Nucl. Phys.}\ }\textbf {\bibinfo {volume} {102}},\
  \bibinfo {pages} {48} (\bibinfo {year} {2018})},\ \Eprint
  {http://arxiv.org/abs/1804.09678} {arXiv:1804.09678 [hep-ph]} \BibitemShut
  {NoStop}%
\bibitem [{\citenamefont {de~Salas}\ \emph {et~al.}(2020)\citenamefont
  {de~Salas}, \citenamefont {Forero}, \citenamefont {Gariazzo}, \citenamefont
  {Mart{\'\i}nez-Mirav{\'e}}, \citenamefont {Mena}, \citenamefont {Ternes},
  \citenamefont {T{\'o}rtola},\ and\ \citenamefont {Valle}}]{deSalas:2020pgw}%
  \BibitemOpen
  \bibfield  {author} {\bibinfo {author} {\bibfnamefont {P.}~\bibnamefont
  {de~Salas}}, \bibinfo {author} {\bibfnamefont {D.}~\bibnamefont {Forero}},
  \bibinfo {author} {\bibfnamefont {S.}~\bibnamefont {Gariazzo}}, \bibinfo
  {author} {\bibfnamefont {P.}~\bibnamefont {Mart{\'\i}nez-Mirav{\'e}}},
  \bibinfo {author} {\bibfnamefont {O.}~\bibnamefont {Mena}}, \bibinfo {author}
  {\bibfnamefont {C.}~\bibnamefont {Ternes}}, \bibinfo {author} {\bibfnamefont
  {M.}~\bibnamefont {T{\'o}rtola}}, \ and\ \bibinfo {author} {\bibfnamefont
  {J.}~\bibnamefont {Valle}},\ }\href@noop {} {\  (\bibinfo {year} {2020})},\
  \Eprint {http://arxiv.org/abs/2006.11237} {arXiv:2006.11237 [hep-ph]}
  \BibitemShut {NoStop}%
\bibitem [{\citenamefont {Baussan}\ \emph {et~al.}(2014)\citenamefont {Baussan}
  \emph {et~al.}}]{Baussan:2013zcy}%
  \BibitemOpen
  \bibfield  {author} {\bibinfo {author} {\bibfnamefont {E.}~\bibnamefont
  {Baussan}} \emph {et~al.} (\bibinfo {collaboration} {ESSnuSB}),\ }\bibfield
  {booktitle} {\emph {\bibinfo {booktitle} {{Proceedings, 2013 Community Summer
  Study on the Future of U.S.~Particle Physics: Snowmass on the Mississippi
  (CSS2013): Minneapolis, MN, USA, July 29-August 6, 2013}}},\ }\href {\doibase
  10.1016/j.nuclphysb.2014.05.016} {\bibfield  {journal} {\bibinfo  {journal}
  {Nucl. Phys.}\ }\textbf {\bibinfo {volume} {B885}},\ \bibinfo {pages} {127}
  (\bibinfo {year} {2014})},\ \Eprint {http://arxiv.org/abs/1309.7022}
  {arXiv:1309.7022 [hep-ex]} \BibitemShut {NoStop}%
\bibitem [{\citenamefont {Wildner}\ \emph {et~al.}(2016)\citenamefont {Wildner}
  \emph {et~al.}}]{Wildner:2015yaa}%
  \BibitemOpen
  \bibfield  {author} {\bibinfo {author} {\bibfnamefont {E.}~\bibnamefont
  {Wildner}} \emph {et~al.} (\bibinfo {collaboration} {ESSnuSB}),\ }\href
  {\doibase 10.1155/2016/8640493} {\bibfield  {journal} {\bibinfo  {journal}
  {Adv. High Energy Phys.}\ }\textbf {\bibinfo {volume} {2016}},\ \bibinfo
  {pages} {8640493} (\bibinfo {year} {2016})}\BibitemShut {NoStop}%
\bibitem [{\citenamefont {Abe}\ \emph {et~al.}(2015)\citenamefont {Abe} \emph
  {et~al.}}]{Abe:2015zbg}%
  \BibitemOpen
  \bibfield  {author} {\bibinfo {author} {\bibfnamefont {K.}~\bibnamefont
  {Abe}} \emph {et~al.} (\bibinfo {collaboration} {Hyper-Kamiokande
  Proto-Collaboration}),\ }\href {\doibase 10.1093/ptep/ptv061} {\bibfield
  {journal} {\bibinfo  {journal} {PTEP}\ }\textbf {\bibinfo {volume} {2015}},\
  \bibinfo {pages} {053C02} (\bibinfo {year} {2015})},\ \Eprint
  {http://arxiv.org/abs/1502.05199} {arXiv:1502.05199 [hep-ex]} \BibitemShut
  {NoStop}%
\bibitem [{\citenamefont {Acciarri}\ \emph {et~al.}(2015)\citenamefont
  {Acciarri} \emph {et~al.}}]{Acciarri:2015uup}%
  \BibitemOpen
  \bibfield  {author} {\bibinfo {author} {\bibfnamefont {R.}~\bibnamefont
  {Acciarri}} \emph {et~al.} (\bibinfo {collaboration} {DUNE}),\ }\href@noop {}
  {\  (\bibinfo {year} {2015})},\ \Eprint {http://arxiv.org/abs/1512.06148}
  {arXiv:1512.06148 [physics.ins-det]} \BibitemShut {NoStop}%
\bibitem [{\citenamefont {Abi}\ \emph {et~al.}(2020)\citenamefont {Abi} \emph
  {et~al.}}]{Abi:2020evt}%
  \BibitemOpen
  \bibfield  {author} {\bibinfo {author} {\bibfnamefont {B.}~\bibnamefont
  {Abi}} \emph {et~al.} (\bibinfo {collaboration} {DUNE}),\ }\href@noop {} {\
  (\bibinfo {year} {2020})},\ \Eprint {http://arxiv.org/abs/2002.03005}
  {arXiv:2002.03005 [hep-ex]} \BibitemShut {NoStop}%
\bibitem [{\citenamefont {An}\ \emph {et~al.}(2016)\citenamefont {An} \emph
  {et~al.}}]{An:2015jdp}%
  \BibitemOpen
  \bibfield  {author} {\bibinfo {author} {\bibfnamefont {F.}~\bibnamefont {An}}
  \emph {et~al.} (\bibinfo {collaboration} {JUNO}),\ }\href {\doibase
  10.1088/0954-3899/43/3/030401} {\bibfield  {journal} {\bibinfo  {journal} {J.
  Phys.}\ }\textbf {\bibinfo {volume} {G43}},\ \bibinfo {pages} {030401}
  (\bibinfo {year} {2016})},\ \Eprint {http://arxiv.org/abs/1507.05613}
  {arXiv:1507.05613 [physics.ins-det]} \BibitemShut {NoStop}%
\bibitem [{\citenamefont {Djurcic}\ \emph {et~al.}(2015)\citenamefont {Djurcic}
  \emph {et~al.}}]{Djurcic:2015vqa}%
  \BibitemOpen
  \bibfield  {author} {\bibinfo {author} {\bibfnamefont {Z.}~\bibnamefont
  {Djurcic}} \emph {et~al.} (\bibinfo {collaboration} {JUNO}),\ }\href@noop {}
  {\  (\bibinfo {year} {2015})},\ \Eprint {http://arxiv.org/abs/1508.07166}
  {arXiv:1508.07166 [physics.ins-det]} \BibitemShut {NoStop}%
\bibitem [{\citenamefont {Holthausen}\ \emph {et~al.}(2013)\citenamefont
  {Holthausen}, \citenamefont {Lindner},\ and\ \citenamefont
  {Schmidt}}]{Holthausen:2012dk}%
  \BibitemOpen
  \bibfield  {author} {\bibinfo {author} {\bibfnamefont {M.}~\bibnamefont
  {Holthausen}}, \bibinfo {author} {\bibfnamefont {M.}~\bibnamefont {Lindner}},
  \ and\ \bibinfo {author} {\bibfnamefont {M.~A.}\ \bibnamefont {Schmidt}},\
  }\href {\doibase 10.1007/JHEP04(2013)122} {\bibfield  {journal} {\bibinfo
  {journal} {JHEP}\ }\textbf {\bibinfo {volume} {04}},\ \bibinfo {pages} {122}
  (\bibinfo {year} {2013})},\ \Eprint {http://arxiv.org/abs/1211.6953}
  {arXiv:1211.6953 [hep-ph]} \BibitemShut {NoStop}%
\bibitem [{\citenamefont {Antusch}\ \emph {et~al.}(2003)\citenamefont
  {Antusch}, \citenamefont {Kersten}, \citenamefont {Lindner},\ and\
  \citenamefont {Ratz}}]{Antusch:2003kp}%
  \BibitemOpen
  \bibfield  {author} {\bibinfo {author} {\bibfnamefont {S.}~\bibnamefont
  {Antusch}}, \bibinfo {author} {\bibfnamefont {J.}~\bibnamefont {Kersten}},
  \bibinfo {author} {\bibfnamefont {M.}~\bibnamefont {Lindner}}, \ and\
  \bibinfo {author} {\bibfnamefont {M.}~\bibnamefont {Ratz}},\ }\href {\doibase
  10.1016/j.nuclphysb.2003.09.050} {\bibfield  {journal} {\bibinfo  {journal}
  {Nucl. Phys. B}\ }\textbf {\bibinfo {volume} {674}},\ \bibinfo {pages} {401}
  (\bibinfo {year} {2003})},\ \Eprint {http://arxiv.org/abs/hep-ph/0305273}
  {arXiv:hep-ph/0305273} \BibitemShut {NoStop}%
\bibitem [{\citenamefont {Antusch}\ \emph {et~al.}(2005)\citenamefont
  {Antusch}, \citenamefont {Kersten}, \citenamefont {Lindner}, \citenamefont
  {Ratz},\ and\ \citenamefont {Schmidt}}]{Antusch:2005gp}%
  \BibitemOpen
  \bibfield  {author} {\bibinfo {author} {\bibfnamefont {S.}~\bibnamefont
  {Antusch}}, \bibinfo {author} {\bibfnamefont {J.}~\bibnamefont {Kersten}},
  \bibinfo {author} {\bibfnamefont {M.}~\bibnamefont {Lindner}}, \bibinfo
  {author} {\bibfnamefont {M.}~\bibnamefont {Ratz}}, \ and\ \bibinfo {author}
  {\bibfnamefont {M.~A.}\ \bibnamefont {Schmidt}},\ }\href {\doibase
  10.1088/1126-6708/2005/03/024} {\bibfield  {journal} {\bibinfo  {journal}
  {JHEP}\ }\textbf {\bibinfo {volume} {03}},\ \bibinfo {pages} {024} (\bibinfo
  {year} {2005})},\ \Eprint {http://arxiv.org/abs/hep-ph/0501272}
  {arXiv:hep-ph/0501272} \BibitemShut {NoStop}%
\bibitem [{\citenamefont {Ohlsson}\ and\ \citenamefont
  {Zhou}(2014)}]{Ohlsson:2013xva}%
  \BibitemOpen
  \bibfield  {author} {\bibinfo {author} {\bibfnamefont {T.}~\bibnamefont
  {Ohlsson}}\ and\ \bibinfo {author} {\bibfnamefont {S.}~\bibnamefont {Zhou}},\
  }\href {\doibase 10.1038/ncomms6153} {\bibfield  {journal} {\bibinfo
  {journal} {Nature Commun.}\ }\textbf {\bibinfo {volume} {5}},\ \bibinfo
  {pages} {5153} (\bibinfo {year} {2014})},\ \Eprint
  {http://arxiv.org/abs/1311.3846} {arXiv:1311.3846 [hep-ph]} \BibitemShut
  {NoStop}%
\bibitem [{\citenamefont {Zhang}\ and\ \citenamefont
  {Zhou}(2016{\natexlab{a}})}]{Zhang:2016djh}%
  \BibitemOpen
  \bibfield  {author} {\bibinfo {author} {\bibfnamefont {J.}~\bibnamefont
  {Zhang}}\ and\ \bibinfo {author} {\bibfnamefont {S.}~\bibnamefont {Zhou}},\
  }\href {\doibase 10.1007/JHEP08(2016)024} {\bibfield  {journal} {\bibinfo
  {journal} {JHEP}\ }\textbf {\bibinfo {volume} {08}},\ \bibinfo {pages} {024}
  (\bibinfo {year} {2016}{\natexlab{a}})},\ \Eprint
  {http://arxiv.org/abs/1604.03039} {arXiv:1604.03039 [hep-ph]} \BibitemShut
  {NoStop}%
\bibitem [{\citenamefont {Zhang}\ and\ \citenamefont
  {Zhou}(2016{\natexlab{b}})}]{Zhang:2016png}%
  \BibitemOpen
  \bibfield  {author} {\bibinfo {author} {\bibfnamefont {J.}~\bibnamefont
  {Zhang}}\ and\ \bibinfo {author} {\bibfnamefont {S.}~\bibnamefont {Zhou}},\
  }\href {\doibase 10.1007/JHEP09(2016)167} {\bibfield  {journal} {\bibinfo
  {journal} {JHEP}\ }\textbf {\bibinfo {volume} {09}},\ \bibinfo {pages} {167}
  (\bibinfo {year} {2016}{\natexlab{b}})},\ \Eprint
  {http://arxiv.org/abs/1606.09591} {arXiv:1606.09591 [hep-ph]} \BibitemShut
  {NoStop}%
\bibitem [{\citenamefont {Gehrlein}\ \emph {et~al.}(2016)\citenamefont
  {Gehrlein}, \citenamefont {Petcov}, \citenamefont {Spinrath},\ and\
  \citenamefont {Titov}}]{Gehrlein:2016fms}%
  \BibitemOpen
  \bibfield  {author} {\bibinfo {author} {\bibfnamefont {J.}~\bibnamefont
  {Gehrlein}}, \bibinfo {author} {\bibfnamefont {S.~T.}\ \bibnamefont
  {Petcov}}, \bibinfo {author} {\bibfnamefont {M.}~\bibnamefont {Spinrath}}, \
  and\ \bibinfo {author} {\bibfnamefont {A.~V.}\ \bibnamefont {Titov}},\ }\href
  {\doibase 10.1007/JHEP11(2016)146} {\bibfield  {journal} {\bibinfo  {journal}
  {JHEP}\ }\textbf {\bibinfo {volume} {11}},\ \bibinfo {pages} {146} (\bibinfo
  {year} {2016})},\ \Eprint {http://arxiv.org/abs/1608.08409} {arXiv:1608.08409
  [hep-ph]} \BibitemShut {NoStop}%
\bibitem [{\citenamefont {Huber}\ \emph {et~al.}(2005)\citenamefont {Huber},
  \citenamefont {Lindner},\ and\ \citenamefont {Winter}}]{Huber:2004ka}%
  \BibitemOpen
  \bibfield  {author} {\bibinfo {author} {\bibfnamefont {P.}~\bibnamefont
  {Huber}}, \bibinfo {author} {\bibfnamefont {M.}~\bibnamefont {Lindner}}, \
  and\ \bibinfo {author} {\bibfnamefont {W.}~\bibnamefont {Winter}},\ }\href
  {\doibase 10.1016/j.cpc.2005.01.003} {\bibfield  {journal} {\bibinfo
  {journal} {Comput. Phys. Commun.}\ }\textbf {\bibinfo {volume} {167}},\
  \bibinfo {pages} {195} (\bibinfo {year} {2005})},\ \Eprint
  {http://arxiv.org/abs/hep-ph/0407333} {arXiv:hep-ph/0407333 [hep-ph]}
  \BibitemShut {NoStop}%
\bibitem [{\citenamefont {Huber}\ \emph {et~al.}(2007)\citenamefont {Huber},
  \citenamefont {Kopp}, \citenamefont {Lindner}, \citenamefont {Rolinec},\ and\
  \citenamefont {Winter}}]{Huber:2007ji}%
  \BibitemOpen
  \bibfield  {author} {\bibinfo {author} {\bibfnamefont {P.}~\bibnamefont
  {Huber}}, \bibinfo {author} {\bibfnamefont {J.}~\bibnamefont {Kopp}},
  \bibinfo {author} {\bibfnamefont {M.}~\bibnamefont {Lindner}}, \bibinfo
  {author} {\bibfnamefont {M.}~\bibnamefont {Rolinec}}, \ and\ \bibinfo
  {author} {\bibfnamefont {W.}~\bibnamefont {Winter}},\ }\href {\doibase
  10.1016/j.cpc.2007.05.004} {\bibfield  {journal} {\bibinfo  {journal}
  {Comput. Phys. Commun.}\ }\textbf {\bibinfo {volume} {177}},\ \bibinfo
  {pages} {432} (\bibinfo {year} {2007})},\ \Eprint
  {http://arxiv.org/abs/hep-ph/0701187} {arXiv:hep-ph/0701187 [hep-ph]}
  \BibitemShut {NoStop}%
\bibitem [{\citenamefont {Ghosh}\ and\ \citenamefont
  {Ohlsson}(2020)}]{Ghosh:2019sfi}%
  \BibitemOpen
  \bibfield  {author} {\bibinfo {author} {\bibfnamefont {M.}~\bibnamefont
  {Ghosh}}\ and\ \bibinfo {author} {\bibfnamefont {T.}~\bibnamefont
  {Ohlsson}},\ }\href {\doibase 10.1142/S0217732320500583} {\bibfield
  {journal} {\bibinfo  {journal} {Mod. Phys. Lett. A}\ }\textbf {\bibinfo
  {volume} {35}},\ \bibinfo {pages} {2050058} (\bibinfo {year} {2020})},\
  \Eprint {http://arxiv.org/abs/1906.05779} {arXiv:1906.05779 [hep-ph]}
  \BibitemShut {NoStop}%
\bibitem [{\citenamefont {Blennow}\ \emph
  {et~al.}(2020{\natexlab{b}})\citenamefont {Blennow}, \citenamefont
  {Fernandez-Martinez}, \citenamefont {Ota},\ and\ \citenamefont
  {Rosauro-Alcaraz}}]{Blennow:2019bvl}%
  \BibitemOpen
  \bibfield  {author} {\bibinfo {author} {\bibfnamefont {M.}~\bibnamefont
  {Blennow}}, \bibinfo {author} {\bibfnamefont {E.}~\bibnamefont
  {Fernandez-Martinez}}, \bibinfo {author} {\bibfnamefont {T.}~\bibnamefont
  {Ota}}, \ and\ \bibinfo {author} {\bibfnamefont {S.}~\bibnamefont
  {Rosauro-Alcaraz}},\ }\href {\doibase 10.1140/epjc/s10052-020-7761-9}
  {\bibfield  {journal} {\bibinfo  {journal} {Eur. Phys. J.}\ }\textbf
  {\bibinfo {volume} {C80}},\ \bibinfo {pages} {190} (\bibinfo {year}
  {2020}{\natexlab{b}})},\ \Eprint {http://arxiv.org/abs/1912.04309}
  {arXiv:1912.04309 [hep-ph]} \BibitemShut {NoStop}%
\bibitem [{\citenamefont {Ghosh}\ \emph {et~al.}(2020)\citenamefont {Ghosh},
  \citenamefont {Ohlsson},\ and\ \citenamefont
  {Rosauro-Alcaraz}}]{Ghosh:2019zvl}%
  \BibitemOpen
  \bibfield  {author} {\bibinfo {author} {\bibfnamefont {M.}~\bibnamefont
  {Ghosh}}, \bibinfo {author} {\bibfnamefont {T.}~\bibnamefont {Ohlsson}}, \
  and\ \bibinfo {author} {\bibfnamefont {S.}~\bibnamefont {Rosauro-Alcaraz}},\
  }\href {\doibase 10.1007/JHEP03(2020)026} {\bibfield  {journal} {\bibinfo
  {journal} {JHEP}\ }\textbf {\bibinfo {volume} {03}},\ \bibinfo {pages} {026}
  (\bibinfo {year} {2020})},\ \Eprint {http://arxiv.org/abs/1912.10010}
  {arXiv:1912.10010 [hep-ph]} \BibitemShut {NoStop}%
\bibitem [{\citenamefont {Abe}\ \emph {et~al.}(2018)\citenamefont {Abe} \emph
  {et~al.}}]{Abe:2016ero}%
  \BibitemOpen
  \bibfield  {author} {\bibinfo {author} {\bibfnamefont {K.}~\bibnamefont
  {Abe}} \emph {et~al.} (\bibinfo {collaboration} {Hyper-Kamiokande}),\ }\href
  {\doibase 10.1093/ptep/pty044} {\bibfield  {journal} {\bibinfo  {journal}
  {PTEP}\ }\textbf {\bibinfo {volume} {2018}},\ \bibinfo {pages} {063C01}
  (\bibinfo {year} {2018})},\ \Eprint {http://arxiv.org/abs/1611.06118}
  {arXiv:1611.06118 [hep-ex]} \BibitemShut {NoStop}%
\bibitem [{\citenamefont {Forero}\ \emph {et~al.}(2017)\citenamefont {Forero},
  \citenamefont {Hawkins},\ and\ \citenamefont {Huber}}]{Forero:2017vrg}%
  \BibitemOpen
  \bibfield  {author} {\bibinfo {author} {\bibfnamefont {D.~V.}\ \bibnamefont
  {Forero}}, \bibinfo {author} {\bibfnamefont {R.}~\bibnamefont {Hawkins}}, \
  and\ \bibinfo {author} {\bibfnamefont {P.}~\bibnamefont {Huber}},\
  }\href@noop {} {\  (\bibinfo {year} {2017})},\ \Eprint
  {http://arxiv.org/abs/1710.07378} {arXiv:1710.07378 [hep-ph]} \BibitemShut
  {NoStop}%
\bibitem [{\citenamefont {Huber}\ \emph {et~al.}(2020)\citenamefont {Huber},
  \citenamefont {Minakata},\ and\ \citenamefont {Pestes}}]{Huber:2019frh}%
  \BibitemOpen
  \bibfield  {author} {\bibinfo {author} {\bibfnamefont {P.}~\bibnamefont
  {Huber}}, \bibinfo {author} {\bibfnamefont {H.}~\bibnamefont {Minakata}}, \
  and\ \bibinfo {author} {\bibfnamefont {R.}~\bibnamefont {Pestes}},\ }\href
  {\doibase 10.1103/PhysRevD.101.093002} {\bibfield  {journal} {\bibinfo
  {journal} {Phys. Rev.}\ }\textbf {\bibinfo {volume} {D101}},\ \bibinfo
  {pages} {093002} (\bibinfo {year} {2020})},\ \Eprint
  {http://arxiv.org/abs/1912.02426} {arXiv:1912.02426 [hep-ph]} \BibitemShut
  {NoStop}%
\bibitem [{\citenamefont {Cowan}\ \emph {et~al.}(2011)\citenamefont {Cowan},
  \citenamefont {Cranmer}, \citenamefont {Gross},\ and\ \citenamefont
  {Vitells}}]{Cowan:2010js}%
  \BibitemOpen
  \bibfield  {author} {\bibinfo {author} {\bibfnamefont {G.}~\bibnamefont
  {Cowan}}, \bibinfo {author} {\bibfnamefont {K.}~\bibnamefont {Cranmer}},
  \bibinfo {author} {\bibfnamefont {E.}~\bibnamefont {Gross}}, \ and\ \bibinfo
  {author} {\bibfnamefont {O.}~\bibnamefont {Vitells}},\ }\href {\doibase
  10.1140/epjc/s10052-011-1554-0, 10.1140/epjc/s10052-013-2501-z} {\bibfield
  {journal} {\bibinfo  {journal} {Eur. Phys. J.}\ }\textbf {\bibinfo {volume}
  {C71}},\ \bibinfo {pages} {1554} (\bibinfo {year} {2011})},\ \bibinfo {note}
  {[Erratum: Eur. Phys. J. {\bf C73}, 2501 (2013)]},\ \Eprint
  {http://arxiv.org/abs/1007.1727} {arXiv:1007.1727 [physics.data-an]}
  \BibitemShut {NoStop}%
\bibitem [{\citenamefont {Wilks}(1938)}]{Wilks:1938dza}%
  \BibitemOpen
  \bibfield  {author} {\bibinfo {author} {\bibfnamefont {S.}~\bibnamefont
  {Wilks}},\ }\href {\doibase 10.1214/aoms/1177732360} {\bibfield  {journal}
  {\bibinfo  {journal} {Ann. Math. Statist.}\ }\textbf {\bibinfo {volume}
  {9}},\ \bibinfo {pages} {60} (\bibinfo {year} {1938})}\BibitemShut {NoStop}%
\bibitem [{\citenamefont {Barger}\ \emph {et~al.}(2002)\citenamefont {Barger},
  \citenamefont {Marfatia},\ and\ \citenamefont {Whisnant}}]{Barger:2001yr}%
  \BibitemOpen
  \bibfield  {author} {\bibinfo {author} {\bibfnamefont {V.}~\bibnamefont
  {Barger}}, \bibinfo {author} {\bibfnamefont {D.}~\bibnamefont {Marfatia}}, \
  and\ \bibinfo {author} {\bibfnamefont {K.}~\bibnamefont {Whisnant}},\ }\href
  {\doibase 10.1103/PhysRevD.65.073023} {\bibfield  {journal} {\bibinfo
  {journal} {Phys. Rev.}\ }\textbf {\bibinfo {volume} {D65}},\ \bibinfo {pages}
  {073023} (\bibinfo {year} {2002})},\ \Eprint
  {http://arxiv.org/abs/hep-ph/0112119} {arXiv:hep-ph/0112119} \BibitemShut
  {NoStop}%
\bibitem [{\citenamefont {Ghosh}\ \emph {et~al.}(2016)\citenamefont {Ghosh},
  \citenamefont {Ghoshal}, \citenamefont {Goswami}, \citenamefont {Nath},\ and\
  \citenamefont {Raut}}]{Ghosh:2015ena}%
  \BibitemOpen
  \bibfield  {author} {\bibinfo {author} {\bibfnamefont {M.}~\bibnamefont
  {Ghosh}}, \bibinfo {author} {\bibfnamefont {P.}~\bibnamefont {Ghoshal}},
  \bibinfo {author} {\bibfnamefont {S.}~\bibnamefont {Goswami}}, \bibinfo
  {author} {\bibfnamefont {N.}~\bibnamefont {Nath}}, \ and\ \bibinfo {author}
  {\bibfnamefont {S.~K.}\ \bibnamefont {Raut}},\ }\href {\doibase
  10.1103/PhysRevD.93.013013} {\bibfield  {journal} {\bibinfo  {journal} {Phys.
  Rev.}\ }\textbf {\bibinfo {volume} {D93}},\ \bibinfo {pages} {013013}
  (\bibinfo {year} {2016})},\ \Eprint {http://arxiv.org/abs/1504.06283}
  {arXiv:1504.06283 [hep-ph]} \BibitemShut {NoStop}%
\bibitem [{\citenamefont {Akhmedov}\ \emph {et~al.}(2004)\citenamefont
  {Akhmedov}, \citenamefont {Johansson}, \citenamefont {Lindner}, \citenamefont
  {Ohlsson},\ and\ \citenamefont {Schwetz}}]{Akhmedov:2004ny}%
  \BibitemOpen
  \bibfield  {author} {\bibinfo {author} {\bibfnamefont {E.~K.}\ \bibnamefont
  {Akhmedov}}, \bibinfo {author} {\bibfnamefont {R.}~\bibnamefont {Johansson}},
  \bibinfo {author} {\bibfnamefont {M.}~\bibnamefont {Lindner}}, \bibinfo
  {author} {\bibfnamefont {T.}~\bibnamefont {Ohlsson}}, \ and\ \bibinfo
  {author} {\bibfnamefont {T.}~\bibnamefont {Schwetz}},\ }\href {\doibase
  10.1088/1126-6708/2004/04/078} {\bibfield  {journal} {\bibinfo  {journal}
  {JHEP}\ }\textbf {\bibinfo {volume} {04}},\ \bibinfo {pages} {078} (\bibinfo
  {year} {2004})},\ \Eprint {http://arxiv.org/abs/hep-ph/0402175}
  {arXiv:hep-ph/0402175} \BibitemShut {NoStop}%
\bibitem [{\citenamefont {Gandhi}\ \emph {et~al.}(2007)\citenamefont {Gandhi},
  \citenamefont {Ghoshal}, \citenamefont {Goswami}, \citenamefont {Mehta},
  \citenamefont {Sankar},\ and\ \citenamefont {Shalgar}}]{Gandhi:2007td}%
  \BibitemOpen
  \bibfield  {author} {\bibinfo {author} {\bibfnamefont {R.}~\bibnamefont
  {Gandhi}}, \bibinfo {author} {\bibfnamefont {P.}~\bibnamefont {Ghoshal}},
  \bibinfo {author} {\bibfnamefont {S.}~\bibnamefont {Goswami}}, \bibinfo
  {author} {\bibfnamefont {P.}~\bibnamefont {Mehta}}, \bibinfo {author}
  {\bibfnamefont {S.}~\bibnamefont {Sankar}}, \ and\ \bibinfo {author}
  {\bibfnamefont {S.}~\bibnamefont {Shalgar}},\ }\href {\doibase
  10.1103/PhysRevD.76.073012} {\bibfield  {journal} {\bibinfo  {journal} {Phys.
  Rev.}\ }\textbf {\bibinfo {volume} {D76}},\ \bibinfo {pages} {073012}
  (\bibinfo {year} {2007})},\ \Eprint {http://arxiv.org/abs/0707.1723}
  {arXiv:0707.1723 [hep-ph]} \BibitemShut {NoStop}%
\bibitem [{\citenamefont {Choubey}\ and\ \citenamefont
  {Roy}(2004)}]{Choubey:2003yp}%
  \BibitemOpen
  \bibfield  {author} {\bibinfo {author} {\bibfnamefont {S.}~\bibnamefont
  {Choubey}}\ and\ \bibinfo {author} {\bibfnamefont {P.}~\bibnamefont {Roy}},\
  }\href {\doibase 10.1103/PhysRevLett.93.021803} {\bibfield  {journal}
  {\bibinfo  {journal} {Phys. Rev. Lett.}\ }\textbf {\bibinfo {volume} {93}},\
  \bibinfo {pages} {021803} (\bibinfo {year} {2004})},\ \Eprint
  {http://arxiv.org/abs/hep-ph/0310316} {arXiv:hep-ph/0310316} \BibitemShut
  {NoStop}%
\end{thebibliography}%

\end{document}